\documentclass[sigconf]{acmart}

\usepackage{algorithm}
\usepackage{algorithmic}
\usepackage{xspace}
\usepackage{multirow}
\usepackage{graphicx}
\usepackage{subcaption} 
\usepackage{float} 
\usepackage{fixltx2e} 
\usepackage{cleveref}
\usepackage{enumitem}
\usepackage{makecell}
\usepackage{thm-restate}
\usepackage{fixfoot}

\newcommand{\aris}[1]{\noindent\textcolor{orange}{\textbf{[Aris:]}~{#1}}\xspace}

\newcommand{\honglian}[1]{\noindent\textcolor{magenta}{\textbf{[Honglian:]}~{#1}}\xspace}

\numberwithin{equation}{section}

\newtheorem{definition}{Definition}
\newtheorem{problem}{Problem}

\newtheorem{prop}{Proposition}

\newcommand{\checkedagtext}[1]{{\color{black}{#1}}}

\newcommand{\para}[1]{{\noindent{\textbf{#1}}}}

\newcommand{\reals}{\ensuremath{\mathbb{R}}\xspace}
\newcommand{\bigO}{\ensuremath{\mathcal{O}}\xspace}
\newcommand{\matr}[1]{\ensuremath{\mathbf{{#1}}}\xspace}
\newcommand{\vect}[1]{\ensuremath{\mathbf{{#1}}}\xspace}
\newcommand{\transmatrix}
{\ensuremath{\matr{P}}\xspace}

\newcommand{\newrestartvec}{\ensuremath{\hat{\matr{v}}}\xspace}
\newcommand{\newtransmatrix}{\ensuremath{\hat{\matr{P}}}\xspace}
\newcommand{\newtransmatrixopt}{\ensuremath{\matr{P}^{*}}\xspace}
\newcommand{\Amatrix}{\ensuremath{\matr{A}}\xspace}
\newcommand{\invUmatrix}{\ensuremath{\matr{U}^{-1}}\xspace}
\newcommand{\Ematrix}{\ensuremath{\matr{E}}\xspace}
\newcommand{\Hmatrix}{\ensuremath{\matr{H}}\xspace}
\newcommand{\Xmatrix}{\ensuremath{\matr{X}}\xspace}
\newcommand{\Ymatrix}{\ensuremath{\matr{Y}}\xspace}
\newcommand{\Cmatrix}{\ensuremath{\matr{C}}\xspace}
\newcommand{\ubound}{\ensuremath{u}\xspace}
\newcommand{\lbound}{\ensuremath{\ell}\xspace}

\newcommand{\Total}{\ensuremath{\text{Total}}\xspace}
\newcommand{\act}{\ensuremath{\text{Active}}\xspace}
\newcommand{\clip}{\ensuremath{\text{clip}}\xspace}
\newcommand{\lambdaopt}{\ensuremath{\lambda^{*}}\xspace}
\newcommand{\vectalpha}{\ensuremath{\boldsymbol{\alpha}}\xspace}

\newcommand{\vects}{\ensuremath{\vect{s}}\xspace}
\newcommand{\vectsopt}{\ensuremath{\vect{s}^{*}}\xspace}
\newcommand{\vectt}{\ensuremath{\vect{t}}\xspace}

\newcommand{\vectl}{\ensuremath{\vect{\ell}}\xspace}
\newcommand{\vectu}{\ensuremath{\vect{u}}\xspace}

\newcommand{\vectw}{\ensuremath{\vect{w}}\xspace}
\newcommand{\vectz}{\ensuremath{\vect{z}}\xspace}
\newcommand{\vectx}{\ensuremath{\vect{x}}\xspace}
\newcommand{\vecty}{\ensuremath{\vect{y}}\xspace}
\newcommand{\vectv}{\ensuremath{\vect{v}}\xspace}
\newcommand{\vectlambda}{\ensuremath{\boldsymbol{\lambda}}\xspace}
\newcommand{\vecte}{\ensuremath{\vect{e}}\xspace}
\newcommand{\prvector}{\ensuremath{\vect{p}}\xspace}
\newcommand{\indvect}[1]{\ensuremath{\vect{1}_{#1}}\xspace} 
\newcommand{\uniformvect}{\ensuremath{\mathbf{1}}\xspace} 

\newcommand{\graph}{\ensuremath{G}\xspace}
\newcommand{\vertices}{\ensuremath{V}\xspace}
\def\group_#1{\vertices_{#1}}
\def\labelof_#1{\ensuremath{\ell}_{#1}}
\newcommand{\edges}{\ensuremath{E}\xspace}
\newcommand{\novertices}{\ensuremath{n}\xspace}
\newcommand{\noedges}{\ensuremath{m}\xspace}

\newcommand{\vertexv}{\ensuremath{v}\xspace}
\newcommand{\unitary}{\ensuremath{\matr{I}}\xspace}
\newcommand{\matrixU}{\ensuremath{\matr{U}}\xspace}

\newcommand{\nogroups}{\ensuremath{K}\xspace}
\newcommand{\agroup}{\ensuremath{k}\xspace}
\newcommand{\grouppr}{\ensuremath{\phi}\xspace}
\newcommand{\groupDist}{\ensuremath{\pmb{\phi}}\xspace}

\newcommand{\ranking}{\ensuremath{\mathbf{r}}\xspace}

\newcommand{\loss}{\ensuremath{L}\xspace}
\newcommand{\lossprev}{\ensuremath{\loss_{\mathit{prev}}}\xspace}
\newcommand{\groupAdapLoss}{\ensuremath{\loss_{\mathit{g}}}\xspace}

\newcommand{\feasiblematrices}{\ensuremath{\mathcal{C}}\xspace}
\newcommand{\resfeasiblematrices}{\ensuremath{\mathcal{C}_R}\xspace}

\newcommand{\boundfactor}{\ensuremath{\delta}\xspace}
\newcommand{\absboundfactor}{\ensuremath{\epsilon}\xspace}

\newcommand{\iter}{\ensuremath{{\mathit{iter}}}\xspace}

\newcommand{\fpr}{\ensuremath{\boldsymbol{\phi}}{\sc\small PR}\xspace}

\newcommand{\resfpr}{{\sc\small R}-{\fpr}\xspace}

\newcommand{\pagerank}{Page\-Rank\xspace}
\newcommand{\phiPageRank}{\groupDist-\pagerank}
\newcommand{\resphiPageRank}{Restricted \phiPageRank}
\newcommand{\gaphiPageRank}{Group-adapted \phiPageRank}
\newcommand{\resgaphiPageRank}{Restricted group-adapted \phiPageRank}

\newcommand{\yapprox}{{\texttt{yApprox}}\xspace}
\newcommand{\project}{{\texttt{Project}}\xspace}
\newcommand{\searchlambda}{{\texttt{Root-Finding}}\xspace}

\newcommand{\fairgd}{{\texttt{FairGD}}\xspace}

\newcommand{\fairgdrone}{\ensuremath{\fairgd}($0.1$,~$0.1$)\xspace}
\newcommand{\fairgdrtwo}{\ensuremath{\fairgd}($0.5$,~$0.1$)\xspace}
\newcommand{\crossgd}{{\texttt{AdaptGD}}\xspace}
\newcommand{\crossgdrone}{\ensuremath{\crossgd}($0.1$,~$0.1$)\xspace}
\newcommand{\crossgdrtwo}{\ensuremath{\crossgd}($0.5$,~$0.1$)\xspace}

\newcommand{\fairwalk}{\textsf{FairWalk}\xspace}
\newcommand{\crosswalk}{\textsf{CrossWalk}\xspace}
\newcommand{\lfpru}{\textsf{LFPR\textsubscript{U}}\xspace}
\newcommand{\lfprn}{\textsf{LFPR\textsubscript{N}}\xspace}

\newcommand{\book}{\textsf{Books}\xspace}
\newcommand{\blogs}{\textsf{Blogs}\xspace}
\newcommand{\twitter}{\textsf{Twitter}\xspace}
\newcommand{\slashdot}{\textsf{Slashdot}\xspace}
\newcommand{\mind}{\textsf{MIND}\xspace}

\DeclareFixedFootnote{\fnlabels}{The original dataset contains 14 labels. We aggregate vertices with the labels \{lifestyle, food and drink, sports, health\} into group 1, and those with the labels \{entertainment, travel, tv, news, finance, movies, autos, music, weather, video\} into group~2.}

\AtBeginDocument{%
  }

\copyrightyear{2026}
\acmYear{2026}
\setcopyright{cc}
\setcctype{by}
\acmConference[WSDM '26]{Proceedings of the Nineteenth ACM International Conference on Web Search and Data Mining}{February 22--26, 2026}{Boise, ID, USA}
\acmBooktitle{Proceedings of the Nineteenth ACM International Conference on Web Search and Data Mining (WSDM '26), February 22--26, 2026, Boise, ID, USA}
\acmPrice{}
\acmDOI{10.1145/3773966.3777991}
\acmISBN{979-8-4007-2292-9/2026/02}

\begin{document}

\title{Fairness-aware PageRank via Edge Reweighting}

\author{Honglian Wang}
\affiliation{%
	\institution{KTH Royal Institute of Technology}
	\institution{Digital Futures}
	\city{Stockholm}
	\country{Sweden}
}
\email{honglian@kth.se}

\author{Haoyun Zhou}
\affiliation{%
	\institution{KTH Royal Institute of Technology}
	\institution{Digital Futures}
	\city{Stockholm}
	\country{Sweden}
}
\email{haoyunz@kth.se}

\author{Aristides Gionis}
\affiliation{%
	\institution{KTH Royal Institute of Technology}
	\institution{Digital Futures}
	\city{Stockholm}
	\country{Sweden}
}
\email{argioni@kth.se}

\begin{abstract}
Link-analysis algorithms, such as \pagerank, are instrumental in understanding the structural dynamics of networks by evaluating the importance of individual vertices based on their connectivity. Recently, with the rising importance of responsible AI, the question of fairness in link-analysis algorithms has gained traction.

In this paper, we present a new approach for incorporating group fairness into the \pagerank algorithm by reweighting the transition probabilities in the underlying transition matrix. We formulate the problem of achieving fair \pagerank by seeking to minimize the fairness loss, which is the difference between the original group-wise \pagerank distribution and a target \pagerank distribution. We further define a group-adapted fairness notion, which accounts for group homophily by considering random walks with group-biased restart for each group.
Since the fairness loss is non-convex, we propose an efficient projected gradient-descent method for computing locally-optimal edge weights. Unlike earlier approaches, we do not recommend adding new edges to the network, nor do we adjust the restart vector. Instead, we keep the topology of the underlying network unchanged and only modify the relative importance of existing edges. We empirically compare our approach with state-of-the-art baselines and demonstrate the efficacy of our method, where very small changes in the transition matrix lead to significant improvement in the fairness of the \pagerank algorithm.
%
\end{abstract}



\received{20 February 2007}
\received[revised]{12 March 2009}
\received[accepted]{5 June 2009}

\maketitle

\section{Introduction}
\label{section:intro}
Machine-learning algorithms and data-driven decision making are playing 
increasingly crucial roles in our lives, 
impacting job markets, healthcare, government, industry, and education. 
As our dependence on automated decisions grows, however, 
so does the concern for fairness in algorithms. 
Ensuring that machine-learning algorithms are reliable
and satisfy strong notions of fairness
has become an issue of societal concern and, 
justifiably, has received significant attention in the literature \citep{fu2022fair,kearns2017fair}.
To date, algorithmic fairness has been studied for primarily supervised~\citep{hardt2016equality,joseph2016fairness,pedreshi2008discrimination,zemel2013learning}
but also unsupervised learning~\citep{bera2019fair,chen2019proportionally,chierichetti2017fair}.
In the context of network and link-analysis algorithms, 
a few different approaches 
have been proposed \citep{khajehnejad_crosswalk_2022,rahman_fairwalk_2019,tsioutsiouliklis_fairness-aware_2021,tsioutsiouliklis_link_2022,zeng_fair_2021}, 
however, the topic has received significantly less attention
than the study of fairness in other settings. 

Link-analysis algorithms, such as the \pagerank family of methods, 
are fundamental tools for analyzing large-scale network data~\citep{gleich_pagerank_2015,bahmani2012pagerank}, 
and have been used in a variety of applications, including 
identifying authoritative vertices~\citep{wang2020personalized}, 
detecting communities~\citep{tabrizi2013personalized}, 
uncovering patterns of influence~\citep{liu2017influence}, and 
mitigating network vulnerabilities~\citep{kay2021identification}. 

\vspace{1mm}
\para{Fairness-aware \pagerank.}
The problem of fairness-aware PageRank~\citep{tsioutsiouliklis_fairness-aware_2021} 
can be broadly formulated as follows.
We are interested in computing the \pagerank scores
on a graph whose vertices are partitioned into \nogroups groups, 
where the partition is defined by, say, a protected attribute. 
For simplicity, let us assume $\nogroups=2$ and refer to the two groups 
of vertices as \emph{red} and \emph{blue}.
The fairness of the \pagerank distribution is then evaluated
based on the total \pagerank scores assigned to \emph{red} and \emph{blue} vertices.
We would like to avoid situations where the \pagerank mass is distributed unevenly (/unfairly) 
between the two groups. 
For example, given a target score tuple 
$\groupDist = (\phi_{\text{red}}, \phi_{\text{blue}})= \checkedagtext{(\phi_{\text{red}}, 1-\phi_{\text{red}})}$, 
we say that \pagerank is $\groupDist$-fair if the total \pagerank score of the
vertices in the \emph{red} group is equal to~$\phi_\mathrm{red}$; 
\checkedagtext{and consequently, the total \pagerank score of the
vertices in the \emph{blue} group is equal to~$1-\phi_{\text{red}}=\phi_\mathrm{blue}$; }


\vspace{1mm}
\para{Intervention strategies.}
\checkedagtext{To improve fairness for \pagerank problems,} 
the task is to design an intervention strategy so that the \pagerank
scores of the graph vertices become $\groupDist$-fair, or ``close enough'' to $\groupDist$-fair.
Considering the \pagerank algorithm, 
there are several possible ways to design intervention strategies, including:
$(i)$~modifying the \pagerank restart probability;
$(ii)$~modifying the restart vector;
$(iii)$~adding edges to the graph; and
$(iv)$~modifying the edge weights of the graph in a controlled manner to ensure that the modified graph remains faithful to the original.

In this paper, we focus on intervention strategies of type ($iv$),
namely, achieving fair \pagerank by re\-weighting the graph edges.
Our motivation is that such interventions are 
\emph{intrinsic to the graph structure} and can be implemented with actionable operations. 
For example, edge re\-weighting can be achieved by prioritizing/demoting content from one's connections in a social network. Strategies of type ($iii$) involving edge additions or deletions are also intrinsic to the graph structure but tend to be less practical. These strategies rely on link recommendations and have to account for the likelihood of accepting the recommended links, which introduces uncertainty and makes them challenging to implement.

Strategies of type ($i$) and ($ii$), i.e., modifying the \pagerank restart probability or restart vector are \emph{extrinsic to the graph structure} and rely on tuning hyper-parameters of the \pagerank algorithm. In many cases, it is not meaningful, nor actionable, to modify those hyper-parameters. For example, in the personalized \pagerank algorithm, the restart vector has a clear interpretation, which is to model the user interests and is not amenable to modification.

\vspace{1mm}
\para{Locally-\,vs.~globally-fair \pagerank.}
Existing papers 
\citep{khajehnejad_crosswalk_2022,rahman_fairwalk_2019,tsioutsiouliklis_fairness-aware_2021} 
make the observation 
that if fairness is ensured at each step of the \pagerank random walk,
then fairness is guaranteed for the final \pagerank distribution.
Ensuring fairness at each step (which we call locally-fair \pagerank)
translates to having equal probability of visiting all groups from each vertex. 
While this is a sound approach, 
the requirement of locally-fair \pagerank can be too rigid and may lead to a
disproportionate modification of the graph structure or the edge weights. 
Instead, in this paper, we pose the question of \emph{whether it is possible to 
achieve globally-fair \pagerank without having to impose the strict requirement of 
locally-fair \pagerank}.

As we will see, we answer this question in the~affirmative.

\vspace{1mm}
\para{Our approach.}
We present a new approach to incorporate fairness into the \pagerank algorithm by re\-weighting the transition probabilities of the \pagerank matrix. 
More concretely, we seek to minimize the \pagerank distance to a fair distribution while allowing only for re\-weighting the network edges within pre\-specified limits. 
Our approach handles multiple vertex groups.
Unlike earlier approaches, 
we do not recommend that new edges be added to the network, nor do we adjust the restart vector. 
Instead, we keep the topology of the underlying network unchanged. 

If perfect fairness cannot be achieved,
our formulation aims to find a solution as close as possible to the target fairness~objective.

Our approach provides an actionable environment to improve    \pagerank fairness, 
as the newly computed edge weights 
can be lead to concrete actions, e.g., 
re\-weighting and prioritizing the feed of users in the social-network platform. 

We propose a projected gradient descent method for computing locally-optimal edge re\-weightings and show how it can be implemented efficiently. 
We empirically compare our approach with state-of-the-art baselines
and demonstrate the efficacy of our method, 
where small changes in the transition matrix lead to significant improvement 
in the fairness of the \pagerank algorithm.

\vspace{1mm}
\para{Contributions.}
Our main contributions are:
\begin{itemize}[leftmargin=*]
\item 
{We incorporate group fairness into \pagerank by re\-weighting transition probabilities to achieve solutions as close as possible to the target fairness objective with minimal modifications.}
\item 
We aim for globally-fair \pagerank without the strict requirement of achieving 
local fairness at each vertex.
This can lead to solutions with smaller modifications in the transition matrix.
\item 
{We introduce a group-adapted fairness notion that accounts for group homophily via group-biased restart distributions.
}
\item 
We propose an efficient projected gradient-descent method for finding locally-optimal edge weights.
\item 
We empirically compare our approach with state-of-the-art baselines and demonstrate the efficacy of our method, 
where very small changes in the transition matrix lead to significant improvement in the fairness of the \pagerank algorithm.\end{itemize}
Due to space limitations, proofs and additional material are in the appendices.

\section{Related work}

In recent years, algorithmic fairness in graph mining has emerged as a prominent area of research
\citep{saxena2024fairsna}. This paper focuses on enhancing fairness in the content of the well-known \pagerank algorithm, which has been widely used for ranking, classification, community detection, and more. The objective 
is to allocate \pagerank scores to vertices in a manner that considers both the vertices' group affiliations and the overall structure of the network~\citep{tsioutsiouliklis_fairness-aware_2021,saxena2024fairsna}.

In the standard \pagerank algorithm \citep{gleich_pagerank_2015,brin_anatomy_1998,tabrizi2013personalized}, 
a random-walk process defines the transition from one vertex to its neighbors,
while also with certain probability considering a ``jump'' to one of the vertices defined by a restart vector.
When considering the group association of vertices, defined by a protected attribute, 
analytical studies in the literature~\citep{espin2022inequality,ghoshal2011ranking} 
have unveild the unfairness of \pagerank.
To address such potential group bias, 
FairWalk \citep{rahman_fairwalk_2019} proposes to modify the transition matrix of the random walk
so that the transition probabilities from each vertex to all groups become equal. 
CrossWalk \citep{khajehnejad_crosswalk_2022} further refines this idea by considering 
the local group structure and reweighting edges based on their distance from group boundaries. 
Both FairWalk and CrossWalk adjust the transition matrix using heuristic-based methods.

\citet{tsioutsiouliklis_fairness-aware_2021} introduce a novel fairness metric for \pagerank 
and propose two families of algorithms: $\mathrm{FSPR}$ and $\mathrm{LFPRs}$. 
Methods in the first family, $\mathrm{FSPR}$, retain the original graph structure and edge weights, 
while seeking an optimal jump vector to minimize the disparity between the fair \pagerank vector and 
the original \pagerank vector, subject to fairness constraints. 
On the other hand, similar to FairWalk and CrossWalk, methods in the second family, $\mathrm{LFPRs}$, 
aim to achieve fairness at the individual vertex level and include three variants: 
$\mathrm{LFPR}_N$, $\mathrm{LFPR}_U$, and $\mathrm{LFPR}_P$. 
Each variant reweights the transition matrix in a different way while incorporating a fair jump vector. 
Although the authors suggest that $\mathrm{FSPR}$ and $\mathrm{LFPRs}$ 
can be extended to scenarios involving more than two groups, 
they do not provide details 
for such an extension. 

Additionally, link recommendation has been explored as an alternative approach to promoting fairness in \pagerank. 
In one such paper, \citet{tsioutsiouliklis_link_2022} present analytical formulas 
for calculating the impact of adding edges on fairness.
In a different direction, \citet{zeng_fair_2021} propose a set of de-biasing methods
for fair network representation learning, using techniques based on 
sampling, projections, and graph neural networks.

Other papers consider the problem of infusing fairness into rankings
without focusing specifically into the \pagerank algorithm.
For instance, \citet{krasanakis_applying_2021} propose a model that 
weights training samples to make classifiers fair and 
adapt it to estimate an unbiased personalization that yields fairer vertex ranks. 

In this paper, we propose an edge-reweighting approach to enhance fairness in \pagerank.
Our method differs from previous studies in several ways: 
First, we consider only edge reweighting, which can lead to more actionable operations to achieve fairness.
Second, we focus on global fairness, 
without enforcing the more rigid requirement of fair transitions from each vertex. 
As a result, our method can lead to significantly smaller modification of the transition matrix.
Finally, our approach offers a clear and practical framework for handling effectively
fairness over multiple groups.




\section{Problem definition}


\subsection{Preliminaries}

We consider a directed graph $\graph = (\vertices, \edges)$ with a set $\vertices$ of $\novertices$ vertices and a set $\edges$ of $\noedges$ edges. 
We assume that the set of vertices is partitioned into \nogroups groups $\{\group_1 \cdots \group_{\nogroups}\}$. 
Let $\labelof_{v}$ indicate the group that a vertex $\vertexv \in \vertices$ belongs to.

The \pagerank algorithm takes \graph as input, and produces a \pagerank vector \prvector that indicates the relative importance of each vertex in the graph~\citep{brin_anatomy_1998}.
Despite the existence of multiple equivalent formulations, the \pagerank algorithm is fully characterized by three parameters: the transition matrix~\transmatrix, the restart (or jump) vector~\vectv, and 
the restart probability~$\gamma$. 
%
The \pagerank vector~\prvector is the solution to the equation
$    \prvector^{\mathsf{T}} = (1 - \gamma) \prvector^{\mathsf{T}} \transmatrix + \gamma \vectv^{\mathsf{T}}.$

Here, the transition matrix \transmatrix is row stochastic, derived by normalizing each row of the adjacency matrix of $\graph$. The restart vector~\vectv specifies a distribution over all vertices, indicating where a random walk might restart. The parameter $\gamma$ represents the probability that the random walk will restart at any given step. In the case of directed graph, there may exist sink vertices that have no outgoing edges, so their rows in the transition matrix would be all zeros.
To handle sink vertices, we replace the corresponding zero rows of \transmatrix with the restart vector \vectv.

By solving 
the  equation of \pagerank definition, 
we obtain the \pagerank vector as 
$\prvector^{\mathsf{T}} = \gamma \vectv^{\mathsf{T}} (\unitary - (1 - \gamma) \transmatrix)^{-1}.$ 

\subsection{Fair \pagerank}

Building on the earlier definition of fairness-aware PageRank \citep{tsioutsiouliklis_fairness-aware_2021, tsioutsiouliklis_link_2022}, we consider the vertices of a graph to be partitioned into distinct groups. These groups may represent social categories based on sensitive attributes such as gender, ethnicity, or sexual orientation. We begin by defining the fair PageRank problem, followed by a discussion of various approaches to achieving PageRank fairness, and conclude with a formal definition of the $\groupDist$-\pagerank (\phiPageRank) problem and the Restricted $\groupDist$-\pagerank (\resfpr) problem.

\subsubsection{Group-wise fair \pagerank}
To establish the concept of \pagerank fairness, we assume a \emph{target \pagerank score}, $\grouppr_\agroup$ for each vertex group $\agroup = 1, \ldots, \nogroups$, \checkedagtext{with $\sum_\agroup \grouppr_\agroup = 1$}. This score represents the total \pagerank mass that the algorithm is expected to allocate to the vertices in group~$\vertices_\agroup$. A common approach is to set $\grouppr_\agroup$ proportionally to the size of $\vertices_\agroup$, i.e., $\grouppr_\agroup = |\vertices_\agroup| / \novertices$. However, other choices may be more suitable depending on the specific application. In this work, as in prior studies, we assume that the target \pagerank score $\grouppr_\agroup$ for each group is provided as input to our problem formulation.

Let $\indvect{\agroup} \in \{0, 1\}^{\novertices}$ denote the indicator vector for group $\vertices_\agroup$, where $\indvect{\agroup}[i] = 1$ if vertex $i \in \vertices_\agroup$ and $\indvect{\agroup}[i] = 0$ otherwise. Let \prvector\ be the PageRank vector computed by the PageRank algorithm using a transition matrix \transmatrix, with restart probability~$\gamma$ and restart vector~\vectv. The total PageRank score for group $\vertices_\agroup$ is then given by $\indvect{\agroup}^{\mathsf{T}} \prvector$.

Given an input graph \graph\ and its associated transition matrix \transmatrix, our goal is to find a new transition matrix $\newtransmatrix$, defined on the same input graph \graph, such that the group-wise \pagerank distribution computed over \newtransmatrix\ (defined shortly) closely matches the target group-wise \pagerank distribution $\groupDist = (\grouppr_1, \cdots, \grouppr_\nogroups)$.

\vspace{1mm}
\noindent
\emph{Remark.}  
For any given $\transmatrix$, $\gamma$, and $\vectv$, the total PageRank score for group $\group_\agroup$ is lowered bounded by $\gamma \indvect{\agroup}^{\mathsf{T}} \vectv$ and upper bounded by $1-\gamma + \gamma \indvect{k}^{\mathsf{T}} \vectv$. The lower bound represents the probability that a random walk restarts from group $\agroup$, and it can be derived from the equation $\indvect{\agroup}^{\mathsf{T}} \prvector = (1 - \gamma)  \indvect{\agroup}^{\mathsf{T}}  \transmatrix^{\mathsf{T}} \prvector + \gamma \indvect{\agroup}^{\mathsf{T}} \vectv \geq \gamma \indvect{\agroup}^{\mathsf{T}} \vectv$. The upper bound can be derived from the equation $1 - \sum_{j \neq k} \indvect{j}^{\mathsf{T}} \vectv \leq 1 - \gamma \sum_{j \neq k} \indvect{j}^{\mathsf{T}} \vectv = 1 - \gamma + \gamma \indvect{k}^{\mathsf{T}}$. 
Thus, for fixed $\gamma$ and \vectv, it is clear that it may not always be possible to find a $\newtransmatrix$ that exactly achieves the desired target distribution $\groupDist$.



\subsubsection{Rationale for \pagerank fairness via edge-re\-weighting}
Consider the case where both $\vectv$ and $\gamma$ can be freely adjusted. In this scenario, achieving the target fair distribution $\groupDist$ is straightforward: we can simply set $\vectv[i] = {\grouppr_{\labelof_i}}/{|\group_{\labelof_i}|}$ and $\gamma = 1$. However, adjusting $\vectv$ and $\gamma$ is impractical 
as these parameters are not within our control in actual models; they are merely parameters of \pagerank and do not reflect the actual graph structure. In contrast, modifying the transition matrix is more meaningful; this can be accomplished, for example, by appropriately prioritizing or demoting content from an individual's connections in a social network.

Therefore, we focus solely on adjusting the transition matrix $\transmatrix$ to achieve the desired fairness distribution. To maintain the structure of the input graph \graph, 
we impose the constraint that the new transition matrix $\newtransmatrix$ must satisfy $\newtransmatrix[i, j] = 0$ whenever $\transmatrix[i, j] = 0$. This condition ensures that no new edges are introduced.

\subsubsection{Fairness-loss function and the \phiPageRank problem}
To formalize our problem, we first define the fairness-loss function. It evaluates how much the actual fairness distribution deviates from the target distribution for a given transition matrix $\transmatrix$.

\begin{definition}[\pagerank Problem Instance]
\label{def:pagerank_instance}
{A \emph{PageRank problem instance} consists of a graph \graph with vertices partitioned into groups $\vertices_1, \ldots, \vertices_\nogroups$, a row-stochastic transition matrix \transmatrix, a restart probability $0 < \gamma < 1$, a restart vector $\vectv \in \reals^{\novertices}$ satisfying $\|\vectv\|_1 = 1$, and group-wise target \pagerank scores $\groupDist = (\grouppr_1, \ldots, \grouppr_\nogroups)$ satisfying $\|\groupDist\|_1 = 1$. }
\end{definition}

\begin{definition}[Fairness-loss function] 
\label{definition:fairloss} 
Given a \pagerank problem input as defined in \Cref{def:pagerank_instance}, the {fairness-loss function} is defined~as: 
\begin{equation} 
\loss(\transmatrix, \gamma, \vectv) = \frac{1}{\nogroups} \sum_{\agroup=1}^\nogroups \left(\indvect{\agroup}^{\mathsf{T}} \prvector - \grouppr_\agroup\right)^2, 
\label{eq:loss_function} 
\end{equation} 
where \prvector is the \pagerank vector corresponding to parameters~$(\transmatrix, \gamma, \vectv)$. \end{definition}

The proof of the following claim can be found in Appendix \ref{appendix:omitted-proposition-non-convexity}

\begin{prop}
\label{proposition:non-convexity}
The fairness-loss function $\loss(\transmatrix, \gamma, \vectv)$, 
as defined in \Cref{eq:loss_function}, is not convex with respect to $\transmatrix$.
\end{prop}

With the fairness-loss function defined, we introduce two problems that aim to achieve fair \pagerank by minimizing this loss function through edge re\-weighting.

\begin{problem}[$\groupDist$-\pagerank]
\label{problem:fair_problem}
{Given a \pagerank problem input as defined in \Cref{def:pagerank_instance}}, the goal is to find a new transition matrix $\newtransmatrixopt$ that minimizes the fairness-loss function $\loss(\newtransmatrixopt, \gamma, \vectv)$, i.e.,
\begin{equation}
\newtransmatrixopt = \arg \min_{\newtransmatrix \in \feasiblematrices(\transmatrix)} \loss(\newtransmatrix, \gamma, \vectv),
\label{eq:fair_problem}
\end{equation}
where 
\begin{equation*}
\feasiblematrices(\transmatrix) = \{ \newtransmatrix \in \mathbb{R}_{\geq 0}^{\novertices \times \novertices} \mid \newtransmatrix \uniformvect = \uniformvect, \, \newtransmatrix_{ij} = 0 \text{ if }~ \transmatrix_{ij} = 0 \}
\end{equation*}
is the set of row-stochastic matrices that have non-negative entries, and the edge-reweighting of \graph can take place only on existing edges.
\end{problem}

\begin{problem}[\resphiPageRank]
\label{problem:restricted-fair_problem}
{Given a \pagerank problem input as defined in \Cref{def:pagerank_instance}}, a relative modification bound $\boundfactor \ge 0$ and an absolute modification bound $\absboundfactor \ge 0$, the goal is to find a new transition matrix $\newtransmatrixopt$ that minimizes the fairness-loss function $\loss(\newtransmatrixopt, \gamma, \vectv)$, i.e.,
\begin{equation}
\newtransmatrixopt = \arg \min_{\newtransmatrix \in \resfeasiblematrices(\transmatrix) \cap \feasiblematrices(\transmatrix)} \loss(\newtransmatrix, \gamma, \vectv),
\label{eq:restrict_fair_problem}
\end{equation}
where $\feasiblematrices(\transmatrix)$ is defined as in Problem~\ref{problem:fair_problem}, and
\begin{equation*}
    \resfeasiblematrices(\transmatrix)
    =
    \{ \newtransmatrix \in \mathbb{R}_{\geq 0}^{\novertices \times \novertices} \mid (1 - \boundfactor) \transmatrix[i, j] - \absboundfactor \le \newtransmatrix[i, j] \le (1 + \boundfactor) \transmatrix[i, j] + \absboundfactor\}.
\end{equation*}
\end{problem}

In other words, $\resfeasiblematrices(\transmatrix)$ restricts the entries of the new transition matrix $\newtransmatrix$ within a range of $(1 - \boundfactor)$ and $(1 + \boundfactor)$ of the corresponding entries in the original transition matrix $\transmatrix$, while also imposing an absolute modification bound $\absboundfactor$. The \checkedagtext{relative} bound is necessary because, for very small values in $\transmatrix$, even slight absolute changes can result in large relative variations. 
\textcolor{black}{By choosing $\delta$ and $\epsilon$ such that $(1 - \delta)\transmatrix - \epsilon >0$, we can guarantee no edges are deleted.}

The motivation for defining the \resphiPageRank problem (Problem~\ref{problem:restricted-fair_problem})
is to consider only solutions that make small modifications in the transition matrix $\transmatrix$.

Without loss of generality, in this paper, we study both \phiPageRank and \resphiPageRank with a uniform restart vector $\vectv = \left[ \frac{1}{\novertices},\ldots, \frac{1}{\novertices} \right]$.

\subsection{Group-adapted fair \pagerank}

Real-world networks are mostly homophilic, 
i.e., vertices in the same group are more likely to be connected to each other 
\citep{mcpherson2001birds,tarbush2012homophily}. 
Hence, the total \pagerank scores in group $\vertices_\agroup$ are largely contributed by random walks 
(re)\-starting at vertices in group $\vertices_\agroup$ \cite{tsioutsiouliklis_fairness-aware_2021}. 
Thus, we propose the notion of \emph{group-adapted fairness}, 
which enhances \pagerank by considering random walks that (re)\-start at each group and 
by penalizing the fairness loss for each such~(re)start.

More specifically, let $\vectv_\agroup = \frac{1}{|\vertices_\agroup|} \indvect{\agroup}$ be 
the vector indicating that the random walk restarts uniformly only from group $\agroup$. 
We define the \emph{group-adapted fairness-loss function} as follows:


\begin{definition}[Group-adapted fairness loss]
\label{definition:groupfairloss}
{Given a \pagerank problem input as defined in \Cref{def:pagerank_instance}},
the group-adapted fairness-loss function is defined as 
\begin{equation}
\label{equation:cgfl}
    \groupAdapLoss(\transmatrix, \gamma)  
        = \frac{1}{\nogroups} \sum_{\ell=1}^\nogroups \loss(\transmatrix, \gamma, \vectv_{\ell})
        = \frac{1}{\nogroups^2} \sum_{\ell=1}^\nogroups \sum_{\agroup=1}^\nogroups 
                \left(\indvect{\agroup}^{\mathsf{T}} \prvector_\ell - \grouppr_\agroup\right)^2,
\end{equation}
where $\prvector_\ell$ is the \pagerank vector with respect to parameters 
$(\transmatrix, \gamma, \vectv_{\ell})$.
\end{definition}

In the group-adapted fairness loss, 
we iterate through all groups and sum up the fairness loss 
when restarting the random walk at each group. 
Minimizing the group-adapted fairness-loss function 
leads to the \gaphiPageRank problem.


\begin{problem}[\gaphiPageRank]
\label{problem:cross-fair-loss}
{Given a \pagerank problem input as defined in \Cref{def:pagerank_instance}}, find a new transition matrix $\newtransmatrixopt$ that minimizes the group-adapted fairness-loss function
\begin{equation}
    \newtransmatrixopt = \arg\min_{\newtransmatrix \in \feasiblematrices(\transmatrix)} 
    \groupAdapLoss(\newtransmatrix, \gamma),
\end{equation}
where $\feasiblematrices(\transmatrix)$ is defined as in Problem~\ref{problem:fair_problem}.
\end{problem}

Similarly, we define the restricted 
version of the problem.

\begin{problem}[\resgaphiPageRank]
\label{problem:cross-restricted-fair-loss}
{Given a \pagerank problem input as defined in \Cref{def:pagerank_instance}}, a relative modification bound $\boundfactor \ge 0$ and an absolute modification bound $\absboundfactor \ge 0$, find a new transition matrix $\newtransmatrixopt$ that minimizes the group-adapted fairness-loss function $\groupAdapLoss(\newtransmatrixopt, \gamma)$, i.e.,
\begin{equation*}
    \newtransmatrixopt = \arg\min_{\newtransmatrix \in \resfeasiblematrices(\transmatrix) \cap \feasiblematrices(\transmatrix)} 
                         \groupAdapLoss(\newtransmatrix, \gamma),
\end{equation*}
where $\feasiblematrices(\transmatrix)$ is defined as in Problem~\ref{problem:fair_problem}
and $\resfeasiblematrices(\transmatrix)$ as in Problem~\ref{problem:restricted-fair_problem}.
\end{problem}

\section{Algorithms}

In this section, we present a projected gradient-descent algorithm 
for finding a locally-optimal transition matrix $\newtransmatrix$ for the 
fair \pagerank problems we defined in the previous section. We propose \Cref{alg:fairgd} for problem \Cref{problem:fair_problem} and \Cref{problem:restricted-fair_problem}, and \Cref{alg:crossgd} for problem \Cref{problem:cross-fair-loss} and \Cref{problem:cross-restricted-fair-loss}.

\subsection{Projected gradient descent for \texorpdfstring{$\boldsymbol{\phi}$}{phi}-PageRank}

We first derive the gradient of the fairness-loss function 
$\loss(\transmatrix, \gamma, \vectv)$ 
with respect to the transition matrix \transmatrix.
The proof of the following proposition can be found in Appendix \ref{appendix:omitted-proposition-proof}

\begin{restatable}{prop}{phipagerankgradient}
\label{theorem:gradient-fairness-loss}
The gradient of the fairness-loss function 
$\loss(\transmatrix, \gamma, \vectv)$ 
with respect to the transition matrix \transmatrix is given by
\begin{equation}
\label{equation:gradient}
    \frac{\partial}{\partial \transmatrix} \loss(\transmatrix, \gamma, \vectv) 
         =  \frac{2(1 - \gamma)}{\nogroups} \sum_{k=1}^{\nogroups} 
            \left(\indvect{\agroup}^{\mathsf{T}} \prvector - \grouppr_\agroup\right) \prvector \vecty_\agroup^{\mathsf{T}},   
\end{equation}
where  \prvector is the \pagerank vector with respect to $(\transmatrix, \gamma, \vectv)$, 
and 
$\vecty_\agroup = \matrixU^{-T} \indvect{\agroup} = 
    (\unitary - (1 - \gamma) \transmatrix)^{-1} \indvect{\agroup}$, 
with $\matrixU = \unitary - (1 - \gamma) \transmatrix^{\mathsf{T}}$.
\end{restatable}

\begin{algorithm}[t]
  \caption{\label{alg:fairgd}\fairgd: Projected gradient descent for \phiPageRank and \resphiPageRank}
  \begin{algorithmic}[1]
    \REQUIRE \transmatrix, $\gamma$, \vectv, 
$\groupDist =(\grouppr_1, \ldots, \grouppr_\nogroups)$, group indicator vectors $\indvect{\agroup}$ for $\agroup = \{1, \cdots, \nogroups\}$, learning rate~$\alpha$, iteration exponent~$t_1$ and~$t_2$, convergence criterion $\kappa$, maximum number of iterations $N_{\text{iter}}$, modification bound \boundfactor and \absboundfactor (optional).
    \ENSURE Optimized transition matrix $\newtransmatrix$
    \STATE $\newtransmatrix \gets \transmatrix$, ${\iter} \gets 0$, ${\lossprev} \gets \infty$, $\prvector = \frac{1}{n} \indvect{}$
    \WHILE {$|\loss(\newtransmatrix, \gamma, \vectv) - {\lossprev}| > \kappa $ and $\iter < N_{\text{iter}}$}
      \STATE $\iter \gets \iter + 1$
      \STATE $\prvector \gets \text{iterating } \prvector^{\mathsf{T}} = (1 - \gamma) \prvector^{\mathsf{T}} \newtransmatrix + \gamma \vectv^{\mathsf{T}} \text{ for } t_1 \text{ times}$
      \STATE ${\lossprev} \gets \loss(\newtransmatrix, \gamma, \vectv)$
      \FOR {$\agroup = 1$ to $\nogroups$}
        \STATE $\vecty_{\agroup} \gets$ $\sum_{i=1}^{t_2} (1-\gamma)^i \newtransmatrix^i \indvect{k}$
          \STATE $\newtransmatrix \gets \newtransmatrix - \alpha \frac{2(1 - \gamma)}{\nogroups} (\indvect{\agroup}^{\mathsf{T}} \prvector - \grouppr_\agroup) \prvector \vecty_k^{\mathsf{T}}$
        \ENDFOR
        \STATE $\newtransmatrix \gets {\project}(\newtransmatrix, \delta, \epsilon)$
    \ENDWHILE
    \RETURN $\newtransmatrix$
  \end{algorithmic}
\end{algorithm}

\begin{algorithm}[t]
	\caption{\label{alg:project}\project: projection to feasible set}
	\begin{algorithmic}[1]
		\REQUIRE Transition matrix $\newtransmatrix$, modification bound \boundfactor and \absboundfactor (optional).
		\ENSURE Projected transition matrix $\newtransmatrix$
            \FOR{$i=1, \cdots, n$}
            \STATE $\vects \gets$ nonzero elements of $\newtransmatrix_i$
		\IF{\boundfactor and \absboundfactor are provided} 
        \STATE $\vectl = \max(0, \vects(1-\delta) - \epsilon)$ \label{algoline:simplex_box_start}
        \STATE $\vectu = \min(1, \vects(1+\delta) + \epsilon)$
        \STATE $\lambdaopt \gets \searchlambda(\vects, \vectl, \vectu)$ \label{algoline:root_finding}
        \STATE $s^{*}_j = \clip (\lambdaopt + s_j, \lbound_j, \ubound_j)$ for all $j \in \{1, \cdots, |\vects|\}$ \label{algoline:simplex_box_end}
		\ELSE
		\STATE Rank elements in $\vects$ such that $s_1 \geq \cdots \geq s_{|\vects|}$ \label{algline:simplex_start}
        \STATE $k(\vects) = \displaystyle \max \Bigl\{ k \in \{1, \dots, |\vects|\} \;\big|\; 1 + k\, s_k > \sum_{j \leq k} s_j \Bigr\}$
        \STATE $\tau(\vects) = \frac{1}{k(\vects)} \left(\sum_{j \leq k(\vects)} s_j -1\right)$
        \STATE $s_j^{*} = \max(0,s_j - \tau(\vects))$ for all $j \in \{1, \cdots, |\vects|\}$
        \label{algline:simplex_end}
	\ENDIF
        \STATE non-zeros entries of $\newtransmatrix_i \gets \vectsopt$
        \ENDFOR
	\end{algorithmic}
\end{algorithm}

\noindent
\textbf{Matrix inversion acceleration.}
The calculation of $\vecty_\agroup = \gamma \matrixU^{-T} \indvect{\agroup}$ involves the inverse of the matrix $\matrixU^{\mathsf{T}} = \unitary - (1 - \gamma) \transmatrix$, which is computationally expensive. To avoid computing matrix inversion, we approximation $\matrixU^{-T}$ with the Neumann series
$    \matrixU^{-T} = \sum_{i=0}^{\infty} (1-\gamma)^i \transmatrix^i,$
and substituting into $\vecty_\agroup$ gives
$    \vecty_\agroup = \sum_{i=0}^{\infty}(1-\gamma)^i \transmatrix^i \indvect{\agroup}. $
As the series terms decrease exponentially in mag\-ni\-tude, 
due to the decaying factor $(1-\gamma)$, 
we can approximate $\vecty_\agroup$ efficiently 
by computing only the first few~series terms.
We show that truncating the Neumann series to its first $50$ terms yields a relative error below $0.0003$. The detailed analysis is provided in Appendix \ref{appendix:Neumann series}.


\vspace{1mm}
\noindent
\textbf{Gradient interpretation.}
In Proposition~\ref{theorem:gradient-fairness-loss} we have defined
$\vecty_\agroup = \matrixU^{-T} \indvect{\agroup}$.
The $j$-th coordinate of~$\vecty_\agroup$ is given by
$\vecty_{\agroup}[j] = \indvect{\agroup}^{\mathsf{T}} (\gamma \matrixU^{-1} \mathbf{e}_j)$,
which represents the sum of \pagerank scores for the vertices in the group $\vertices_\agroup$, with the random walk restarting at the unit vector~$\vecte_j$. In other words, the random walk always (re)\-starts from vertex $j$. 

Hence, the gradient of the fairness loss has the following intuitive interpretation:
the $[i, j]$-th entry of the gradient matrix is a weighted sum of the products of 
$\prvector[i]$ and $\vecty_\agroup[j]$. 
The coordinate $\prvector[i]$ represents the frequency with which a random walk, restarting at \vectv, visits vertex~$i$, 
while the coordinate $\vecty_\agroup[j]$ represents the frequency with which a random walk, restarting at vertex~$j$, 
visits vertices in the group~$\vertices_\agroup$. 
Therefore, the $\agroup$-th term of the summation in Eq.~(\ref{equation:gradient})
reflects the weighted contribution of the edge $(i, j)$ 
to the overall \pagerank score of the group~$\vertices_\agroup$.
The weight of the $\agroup$-th term  in Eq.~(\ref{equation:gradient}) 
is proportional to the difference between the total \pagerank scores 
of group $\vertices_\agroup$ and the group-wise target score~$\grouppr_\agroup$.

\vspace{1mm}
\noindent
\textbf{Gradient projection.} 
After each gradient descent step, 
\newtransmatrix needs to be projected back into the feasible area subject to constraints $\feasiblematrices(\transmatrix)$ or $\feasiblematrices(\transmatrix) \cap \resfeasiblematrices(\transmatrix)$.
We perform the projection for each row of $\newtransmatrix$ separately, and since both constraints forbid adding non-existing edges, we solve the projection only for its non-zero \checkedagtext{entries}. Given any row of \newtransmatrix, let \vects be \checkedagtext{the vector comprises the} non-zero \checkedagtext{entries of the corresponding row vector}, and \vectsopt be the vector after projection. We use $N = |
\vects|$ to denote the cardinality of \vects.




We first discuss the constraint
$\feasiblematrices(\transmatrix)$. Projecting~{\vects} \checkedagtext{onto a vector}~{\vectt} \checkedagtext{that lies in the feasible set} $\feasiblematrices(\transmatrix)$ is equivalent to solving the following problem:
\begin{equation*}
    \vectsopt  =\underset {\vectt \in  \mathcal{F}_1}{\arg\min} \left\| \vects - \vectt \right\|_2^2 ,
\end{equation*}
where $\mathcal{F}_{1} = \{ \vectt \in \mathbb{R}^{N} \mid \mathbf{1}^{\mathsf{T}}\vectt = 1, \vectt \geq 0\}$ is a probability simplex. This is a well studied problem~\citep{michelot1986finite, pardalos1990algorithm, duchi2008efficient, condat2016fast}, and the computational complexity for \checkedagtext{finding} \vectsopt is $\bigO(NlogN)$.
In \Cref{alg:project}, lines {\ref{algline:simplex_start}--\ref{algline:simplex_end}}, we use the method detailed by \citet{duchi2008efficient} to perform the projection of $\vects$ after every gradient update step.

We then discuss the constraint $\feasiblematrices(\transmatrix) \cap \resfeasiblematrices(\transmatrix)$. Solving the projection subject to this constraint is equivalent to \checkedagtext{finding} a Euclidean projection of $\vects$ onto the intersection of a probability simplex and a box. Namely,
\begin{equation*}
\label{eq:projection_simplex_plus_box}
	\vectsopt  =\underset {\vectt \in  \mathcal{F}_2}{\arg\min} \left\| \vects - \vectt \right\|_2^2,
\end{equation*}
where $\mathcal{F}_2 = \{ \vectt \in \mathbb{R}^{N} \mid \mathbf{1}^{\mathsf{T}} \vectt = 1,  \lbound_i \leq t_i \leq \ubound_i \ \text{ for all } i \in \{1, \cdots, N\}\}$, with $\lbound_i = \max \{0, (1 - \delta) s_i - \epsilon \}$ and $\ubound_i = \min \{1, (1 + \delta) s_i + \epsilon \}$. This problem has already been studied by \citet{adam2022projections}, and the time complexity for finding $\vectsopt$ is also $\mathcal{O}(NlogN)$.
We provide the details in Algorithm~\ref{alg:project}, lines \ref{algoline:simplex_box_start}--\ref{algoline:simplex_box_end}. Line \ref{algoline:root_finding} invokes Algorithm~\ref{alg:root_finding} (\searchlambda) as a subroutine, and the details are provided in Appendix \ref{appendix:pseudocode}.

To sum up, our method is presented in Algorithms~\ref{alg:fairgd} and~\ref{alg:project}.
Algorithm~\ref{alg:fairgd} (\fairgd) 
is the overall projected gradient-descent scheme that 
iteratively updates the matrix \newtransmatrix until convergence.
Algorithm~\ref{alg:project} (\project)
is the projection function, 
where the transition matrix~\newtransmatrix is projected back to the feasible set after each~iteration. 


Algorithm~\ref{alg:fairgd} is designed to address both the \phiPageRank problem and the \resphiPageRank problem. In the non-restricted case, we refer to the algorithm simply as \fairgd. In the restricted case, parameterized by $(\delta, \epsilon)$, we denote it as $\fairgd(\delta, \epsilon)$.


\subsection{Group-adapted \pagerank}

The \fairgd algorithm can be extended to address both the \gaphiPageRank and \resgaphiPageRank problems by introducing an outer loop that iterates over all groups. This enhanced algorithm, referred to as \crossgd, 
is detailed in~\Cref{alg:crossgd} in
Appendix \ref{appendix:pseudocode}.
For the non-restricted case, we refer to the algorithm as \crossgd. In the restricted case, parameterized by $(\delta, \epsilon)$, we denote it as $\crossgd(\delta, \epsilon)$.

\begin{restatable}{prop}{gaphipagerankgradient}
\label{theorem:adapted-gradient-fairness-loss}
The gradient of the fairness-loss function 
$\groupAdapLoss(\transmatrix, \gamma)$ 
with respect to the transition matrix \transmatrix is given by
\begin{equation}
\begin{aligned}
\label{eq:gradient_cross_fairness_loss}
    \frac{\partial}{\partial \transmatrix} \groupAdapLoss(\transmatrix, \gamma) 
         = \frac{2 (1-\gamma)}{K^2} \sum_{k=1}^{\nogroups} 
         \sum_{\ell=1}^{\nogroups} \left(\indvect{\agroup}^{\mathsf{T}} \prvector_{\ell} - \grouppr_\agroup\right) \left( \prvector_{\ell} \indvect{\agroup}^{\mathsf{T}} \matrixU^{-1} \right) .  
\end{aligned}
\end{equation}
where  $\prvector_{\ell}$ is the \pagerank vector with respect to $(\transmatrix, \gamma, \vectv_{\ell})$.
\end{restatable}

\subsection{Extensions}


{While not explicitly studied in the paper, we note that the proposed fairness loss can be generalized in several ways. First, when fairness scores are specified at the individual-node level, an auxiliary loss term can jointly enforce both group-level and individual-level constraints. Second, for graphs with hierarchical group structures, where each layer defines a partition with associated fairness goals, the overall objective can be formulated as the sum of layer-specific losses, enabling simultaneous satisfaction of fairness requirements across multiple scales. Finally, the restricted fairness losses can incorporate node-level constraints by specifying $\delta$ and $\epsilon$ for each node, with our algorithm remaining applicable.}

\section{Complexity and convergence analysis}

The primary computational cost arises from computing \pagerank{} and approximating $\vecty_k$. 
The \pagerank{} vector is obtained by iterating the update equation $\prvector^\top = (1 - \gamma)\, \prvector^\top \transmatrix + \gamma\, \vectv^\top$
for~$t_1$ iterations, resulting in computational complexity $\mathcal{O}(t_1 n^2)$. 
The computation of $\vecty_k$ involves multiplying a vector by a matrix $t_2$ times. 
Evaluating $\vecty_k$ for all $k \in \{1, \dots, \nogroups\}$ therefore requires $\mathcal{O}(\nogroups t_2 n^2)$ operations. 
Consequently, the overall complexity of \Cref{alg:fairgd} is $\mathcal{O}\big((\nogroups  t_2 + t_1) n^2\big)$.
Since \crossgd{} performs an additional loop over $\nogroups$ groups when computing $\vecty_k$, its complexity becomes $\mathcal{O}\big((\nogroups^2 t_2 + t_1) n^2\big).$


Both loss objective functions, $\loss(\transmatrix,\gamma,\vectv)$ and $\groupAdapLoss(\transmatrix, \gamma)$, are differentiable functions concerning \transmatrix, and their first derivatives are smooth. Since the feasible set $\resfeasiblematrices(\transmatrix) \cap \feasiblematrices(\transmatrix)$ is convex and closed, the proposed projected gradient update method with constant step size converges to a stationary point, which is a local minimum \cite{bertsekas1997nonlinear}. 
We provide detailed analysis in \Cref{section:convergence_L} and~\Cref{section:convergence_Lg}.

\section{Experimental evaluation}



In our experimental evaluation, we study the following questions:

\vspace{1mm}
\noindent \textbf{Q1:} To what extent can the \fairgd and \crossgd algorithms improve the fairness and group-adapted fairness of \pagerank? 
\vspace{1mm}\\ 
\textbf{Q2:} 
How do the \fairgd and \crossgd algorithms compare to the baselines in terms of changes to the transition matrix? 
\vspace{1mm}\\ 
\textbf{Q3:} How does the ranking of the \pagerank score within each group change? Do \fairgd and \crossgd preserve this ranking?

\vspace{1mm}
To answer the first question, we report the fairness loss and group-adapted fairness loss as defined in Equations
(\ref{eq:loss_function}) and (\ref{equation:cgfl}). 
For the second question, we quantify the changes to the transition matrix by the relative change, 
given by
$  \Delta_\transmatrix = {||\newtransmatrix - \transmatrix||_F}/{||\transmatrix||_F}.$
For the third question, we compute the Spearman's rank correlation coefficient 
\citep{spearman1961proof} between the original PageRank ranking and the ranking after re-weighting, 
and report the weighted average $\overline{\rho}$ across all groups 
(weighted by the group size).
A larger $\overline{\rho}$ indicates fewer changes in the PageRank score's ranking.

\subsection{Datasets}

\begin{table}[t]
\centering
  \caption{Datasets used in the experiments. $\groupDist_{O}$ denotes the original group-wise \pagerank scores of the datasets.}
  \label{table:datasets}
  \vspace{-3mm}
  \begin{small}
  \begin{tabular}{lrrrc}
    \toprule
    Dataset & \#Vertices & \#Edges  & Modularity & $\groupDist_{O}$\\
    \midrule
    \book & 105 & 882  & 0.4149 & ( 0.42,0.10, 0.48) \\
    \blogs & 1224 & 19025 & 0.4111& (0.48, 0.52) \\
    \mind & 2269 & 11345  & 0.075 & ( 0.26,0.74)\\
    \twitter & 18470 & 48365  & 0.4755 & (0.42, 0.58)\\
    \slashdot & 82168 & 948464 & 0.2950 & (0.30, 0.36, 0.17,0.17)\\
    \bottomrule
  \end{tabular}
  \end{small}
\end{table}

We evaluate our methods and the baselines on the following publicly-available datasets:
\book,\footnote{\url{http://www-personal.umich.edu/~mejn/netdata/}}
\blogs~\citep{adamic_political_2005},
\mind~\citep{wu2020mind},
\twitter~\citep{nr-sigkdd16}, and
\slashdot~\citep{leskovec_community_2009}.
Dataset statistics are given in \Cref{table:datasets}, 
whereas more information is presented in Appendix \ref{appendix:datasets}.

\begin{figure*}[t] 
    \centering
    \includegraphics[width=\textwidth]{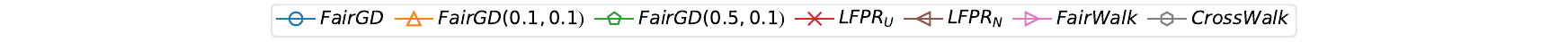}
    
    \begin{subfigure}[b]{0.21\textwidth}
        \centering        \includegraphics[width=\textwidth]{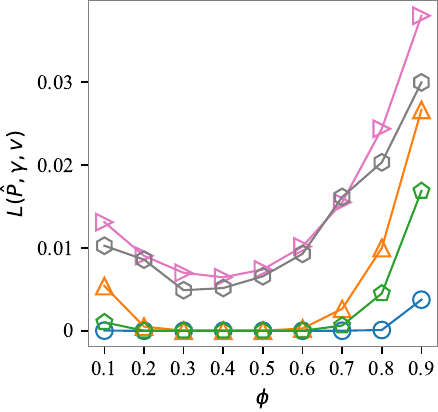} 
        \caption{\book}
\label{fig:loss_book}
    \end{subfigure}
    \hfill
    \begin{subfigure}[b]{0.19\textwidth}
        \centering
\includegraphics[width=\textwidth]{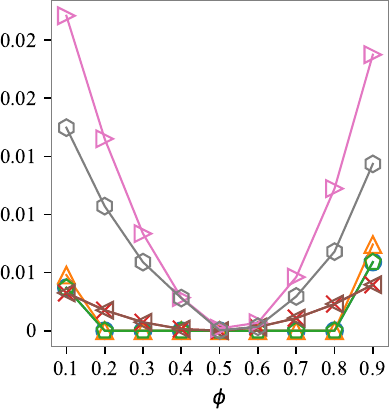}
        \caption{\blogs}
\label{fig:loss_blog}
    \end{subfigure}
    \hfill
        \begin{subfigure}[b]{0.19\textwidth}
        \centering
\includegraphics[width=\textwidth]{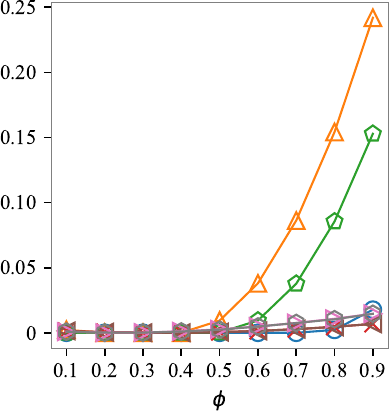}
        \caption{\mind}
\label{fig:loss_mind}
    \end{subfigure}
    \hfill
        \begin{subfigure}[b]{0.19\textwidth}
        \centering
\includegraphics[width=\textwidth]{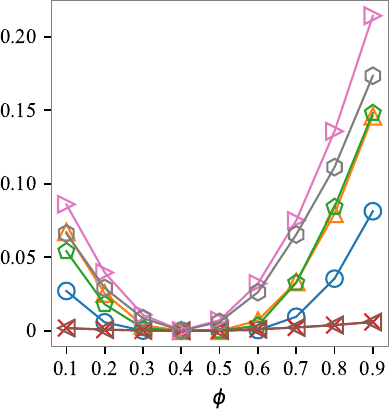}
        \caption{\twitter}
\label{fig:loss_twitter}
    \end{subfigure}
    \hfill
        \begin{subfigure}[b]{0.19\textwidth}
        \centering
\includegraphics[width=\textwidth]{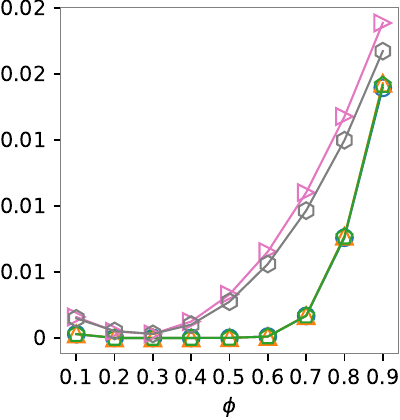}
        \caption{\slashdot}
\label{fig:loss_slashdot}
    \end{subfigure}
    \caption{Fairness loss $\loss(\newtransmatrix,\gamma,\vectv)$ of each method on five datasets for the \phiPageRank problem}
    \label{fig:loss_all_dataset}
\end{figure*}

\begin{figure*}[t] 
    \centering
    \begin{subfigure}[b]{0.21\textwidth}
        \centering        \includegraphics[width=\textwidth]{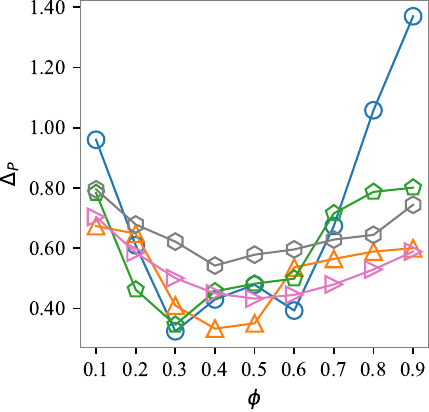} 
        \caption{\book}
\label{fig:change_book}
    \end{subfigure}
    \hfill
    \begin{subfigure}[b]{0.19\textwidth}
        \centering
\includegraphics[width=\textwidth]{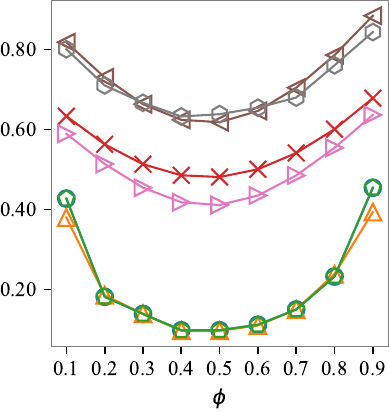}
        \caption{\blogs}
\label{fig:change_blog}
    \end{subfigure}
    \hfill
        \begin{subfigure}[b]{0.19\textwidth}
        \centering
\includegraphics[width=\textwidth]{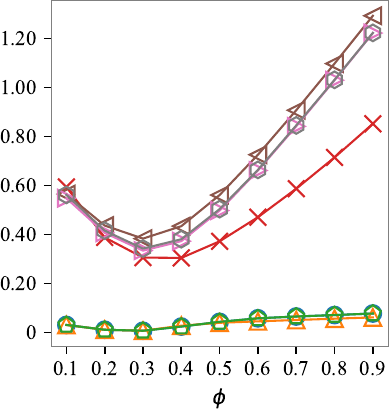}
        \caption{\mind}
\label{fig:change_mind}
    \end{subfigure}
    \hfill
        \begin{subfigure}[b]{0.19\textwidth}
        \centering
\includegraphics[width=\textwidth]{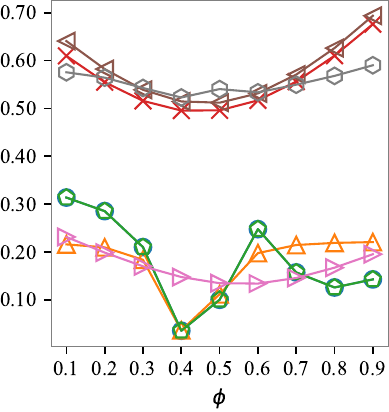}
        \caption{\twitter}
\label{fig:change_twitter}
    \end{subfigure}
    \hfill
        \begin{subfigure}[b]{0.19\textwidth}
        \centering
\includegraphics[width=\textwidth]{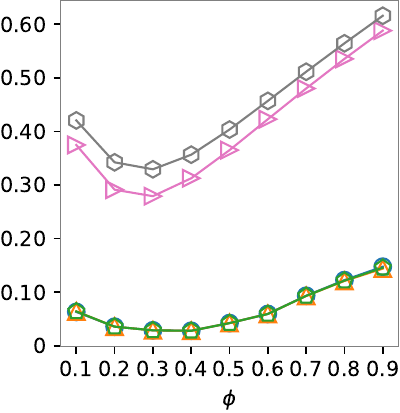}
        \caption{\slashdot}
\label{fig:change_slashdot}
    \end{subfigure}
    \caption{Transition matrix change $\Delta_\transmatrix$ of each method on five datasets for the \phiPageRank problem}
    \label{fig:change_all_dataset}
\end{figure*}


\subsection{Proposed methods and baselines}

The proposed methods are \fairgd and \crossgd. Additionally:

\vspace{1mm}
\noindent 
$\fairgd(\boundfactor,\absboundfactor)$  is the restricted version of \fairgd, 
with relative and absolute intervention bound $\boundfactor$ and $\absboundfactor$ respectively.

\vspace{1mm}
\noindent 
$\crossgd(\boundfactor,\absboundfactor)$ is the restricted version of our \crossgd, 
with relative and absolute intervention bound $\boundfactor$ and $\absboundfactor$ respectively.

\vspace{1mm}
\noindent 
We compare our methods with the following baselines:
\lfpru~\citep{tsioutsiouliklis_fairness-aware_2021},
\lfprn~\citep{tsioutsiouliklis_fairness-aware_2021},
\fairwalk~\citep{rahman_fairwalk_2019}, and
\crosswalk~\citep{khajehnejad_crosswalk_2022}.
More details on the baselines
can be found in Appendix \ref{appendix:baselines}.

\vspace{1mm}
\noindent
\textbf{Remarks.}
Originally, \fairwalk and \crosswalk assign equal weights to each group in the random walk. 
We modify them so that the probability of visiting each group during a random walk is proportional to the target group-wise \pagerank scores. 
These two algorithms do not add edges, 
neither do they change the restart vector.

\lfprn and \lfpru are guaranteed to achieve perfect $\groupDist$-fairness, 
provided the restart vector is $\groupDist$-fair, i.e., $\indvect{\agroup}^{\mathsf{T}} \vectv = \grouppr_\agroup$. \lfpru modifies the graph to be fully connected. We report fairness loss $\loss(\hat{\transmatrix}, \gamma, \vectv)$ and group-adapted fairness loss $\groupAdapLoss(\hat{\transmatrix}, \gamma)$ for these two baselines with the restart vector $\vectv$ set to a uniform vector.

\begin{figure*}[t] 
    \centering
    \includegraphics[width=\textwidth]{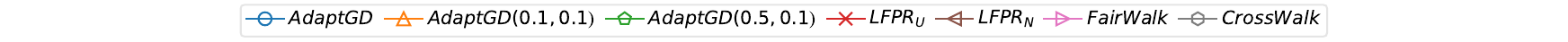}

    \begin{subfigure}[b]{0.21\textwidth}
        \centering        \includegraphics[width=\textwidth]{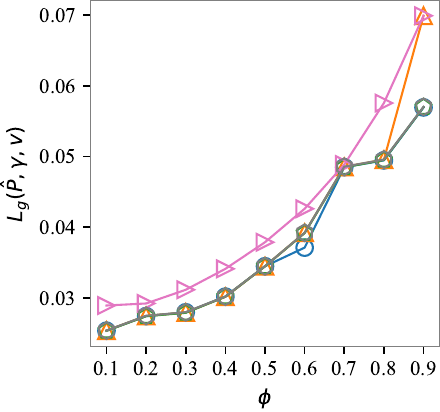} 
        \caption{\book}
\label{fig:adaptloss_book}
    \end{subfigure}
    \hfill
    \begin{subfigure}[b]{0.19\textwidth}
        \centering
\includegraphics[width=\textwidth]{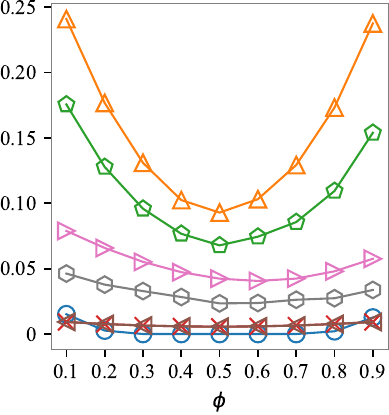}
        \caption{\blogs}
\label{fig:adaptloss_blog}
    \end{subfigure}
    \hfill
        \begin{subfigure}[b]{0.19\textwidth}
        \centering
\includegraphics[width=\textwidth]{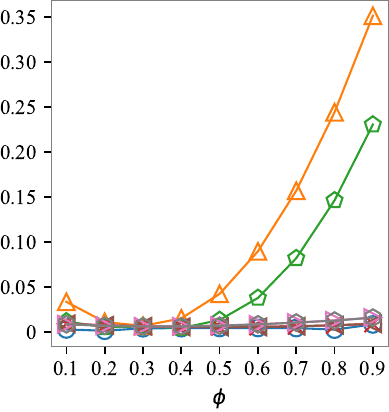}
        \caption{\mind}
\label{fig:adaptloss_mind}
    \end{subfigure}
    \hfill
        \begin{subfigure}[b]{0.19\textwidth}
        \centering
\includegraphics[width=\textwidth]{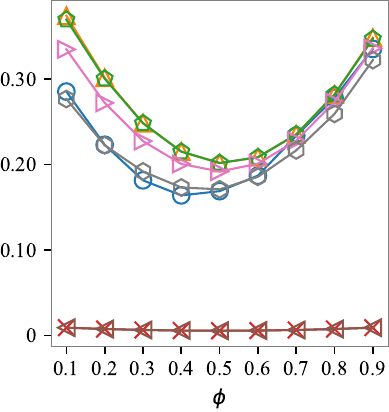}
        \caption{\twitter}
\label{fig:adaptloss_twitter}
    \end{subfigure}
    \hfill
        \begin{subfigure}[b]{0.19\textwidth}
        \centering
\includegraphics[width=\textwidth]{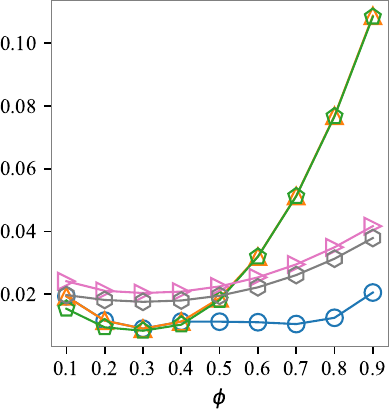}
        \caption{\slashdot}
\label{fig:adaptloss_slashdot}
    \end{subfigure}
    \caption{Fairness loss $\groupAdapLoss(\newtransmatrix,\gamma)$ of each baseline on five datasets for the \gaphiPageRank problem}
    \label{fig:loss_all_dataset_adapted}

     \centering
    \begin{subfigure}[b]{0.21\textwidth}
        \centering        \includegraphics[width=\textwidth]{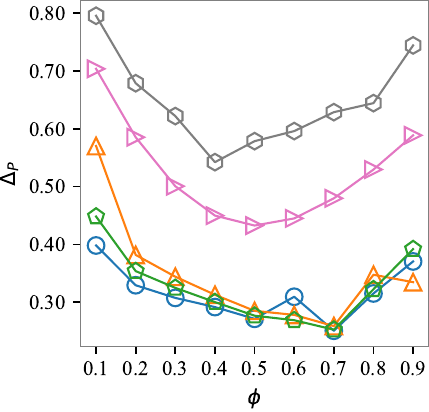} 
        \caption{\book}
\label{fig:adapechange_book}
    \end{subfigure}
    \hfill
    \begin{subfigure}[b]{0.19\textwidth}
        \centering
\includegraphics[width=\textwidth]{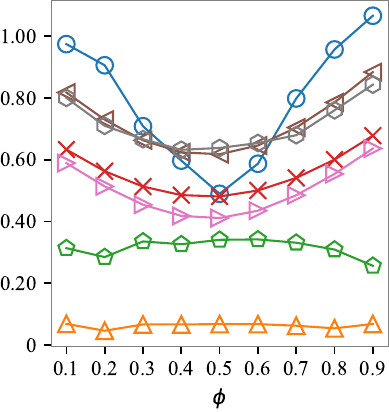}
        \caption{\blogs}
\label{fig:adaptchange_blog}
    \end{subfigure}
    \hfill
        \begin{subfigure}[b]{0.19\textwidth}
        \centering
\includegraphics[width=\textwidth]{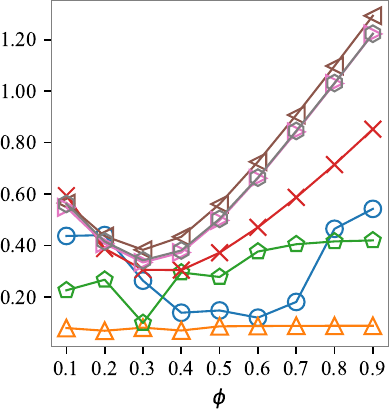}
        \caption{\mind}
\label{fig:adaptchange_mind}
    \end{subfigure}
    \hfill
        \begin{subfigure}[b]{0.19\textwidth}
        \centering
\includegraphics[width=\textwidth]{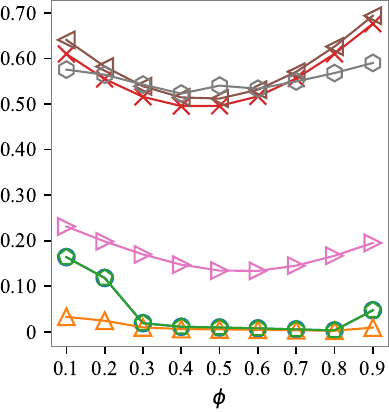}
        \caption{\twitter}
\label{fig:adaptchange_twitter}
    \end{subfigure}
    \hfill
        \begin{subfigure}[b]{0.19\textwidth}
        \centering
\includegraphics[width=\textwidth]{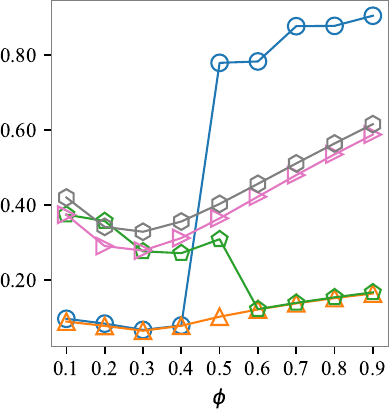}
        \caption{\slashdot}
\label{fig:adaptchange_slashdot}
    \end{subfigure}
    \caption{Transition matrix change $\Delta_\transmatrix$ of each baseline on five datasets for the \gaphiPageRank problem}
    \label{fig:change_all_dataset_adapted}
    
\end{figure*}

   

\subsection{Experimental setting}

\textbf{Parameters $\gamma$ and \vectv}. 
We follow the common assumption of \pagerank algorithm and fix $\gamma = 0.15$. 
We fix $\vectv$ to be the uniform vector.

\vspace{1mm}
\noindent
\textbf{Group indicator definition.} 
For datasets \book, \blogs, \mind, and \twitter, we use node labels as group indicators. For the \slashdot dataset, where node labels are unavailable, we apply the METIS graph partitioning algorithm \cite{karypis_metis_1997} to divide the vertices into groups.

\vspace{1mm}
\noindent
\textbf{(Group-adapted) fairness loss evaluation.} 
For the two-group datasets \blogs, \mind and \twitter datasets, we compare our approach with all baselines.
For the remaining multi-group datasets, we only compare with \fairwalk and \crosswalk as \lfpru and \lfprn do not  extend directly to the multi-group scenario.

\vspace{1mm}
\noindent
\textbf{$\fairgd(\boundfactor,\absboundfactor)$ and $\crossgd(\boundfactor, \absboundfactor)$.}
For \gaphiPageRank and \resgaphiPageRank problems, we set $\boundfactor = \{ 0.1, 0.5\}$ and $\absboundfactor = 0.1$. 
We perform a grid search over a logarithmic\-al\-ly-spaced range of values between $10^{-4}$ and~$10^4$
to find the best learning rate.

\vspace{1mm}
\noindent
\textbf{Group-wise target \pagerank score \groupDist.}
For datasets containing only two groups, we define the target \pagerank distribution as 
$ \groupDist = (\grouppr, 1-\grouppr). $ 
For datasets with multiple groups, we set 
$ \groupDist = \left(\grouppr, \frac{1-\grouppr}{K-1}, \cdots, \frac{1-\grouppr}{K-1}\right). $ 
We vary \(\grouppr\) in the range  $ [0.1, 0.2, \dots, 0.9] $. 

\vspace{1mm}
The code and datasets used in this paper are available.%
\footnote{\url{https://github.com/HongLWang/Fairness-aware-PageRank-via-edge-reweighting}}


\begin{table*}[t]
\centering
\caption{
Average Spearman's coefficient $\overline{\rho}$ between the original and revised \pagerank weight for the \mind dataset}
\vspace{-2mm}
\begin{small}
\begin{tabular}{c|cccccccccc}
\hline
$\phi$ & \fairwalk & \crosswalk & \lfpru & \lfprn &  \fairgd & $\fairgd(0.1, 0.1)$ & $\fairgd(0.5, 0.1)$ & \crossgd & $\crossgd(0.1, 0.1)$ & $\crossgd(0.5, 0.1)$ \\
\hline
$0.1$ & 0.811 & 0.812 & 0.790 & 0.802 & \textbf{0.970} & \underline{0.953} & \textbf{0.970} & 0.817 & 0.853 & 0.830 \\
$0.2$ & 0.830 & 0.823 & 0.808 & 0.801 & \textbf{0.997} & \textbf{0.997} & \textbf{0.997} & 0.810 & \underline{0.855} & 0.817 \\
$0.3$ & 0.818 & 0.808 & 0.817 & 0.810 & \textbf{0.998} & \textbf{0.998} & \textbf{0.998} & 0.797 & \underline{0.835} & 0.815 \\
$0.4$ & 0.828 & 0.795 & 0.811 & 0.795 & \textbf{0.970} & \underline{0.966} & \textbf{0.970} & 0.809 & 0.847 & 0.813 \\
$0.5$ & 0.819 & 0.812 & 0.810 & 0.802 & \textbf{0.970} & 0.909 & \underline{ 0.960} & 0.814 & 0.828 & 0.802 \\
$0.6$ & 0.821 & 0.809 & 0.788 & 0.807 & \textbf{0.966} & 0.888 & \underline{0.903} & 0.814 & 0.846 & 0.825 \\
$0.7$ & 0.816 & 0.805 & 0.789 & 0.786 & \textbf{0.958} & 0.884 & \underline{0.888} & 0.816 & 0.836 & 0.810 \\
$0.8$ & 0.794 & 0.805 & 0.794 & 0.791 & \textbf{0.930} & \underline{0.872} & 0.883 & 0.769 & 0.849 & 0.791 \\
$0.9$ & 0.810 & 0.800 & 0.802 & 0.798 & \textbf{0.884} & 0.857 & \underline{0.867} & 0.760 & 0.833 & 0.801 \\
\hline
\end{tabular}
\end{small}
\label{table:ranking_mind}
\end{table*}

\begin{figure*}[t]
    \centering
    \begin{subfigure}[b]{0.164\textwidth}
        \centering
        \includegraphics[width=\textwidth]{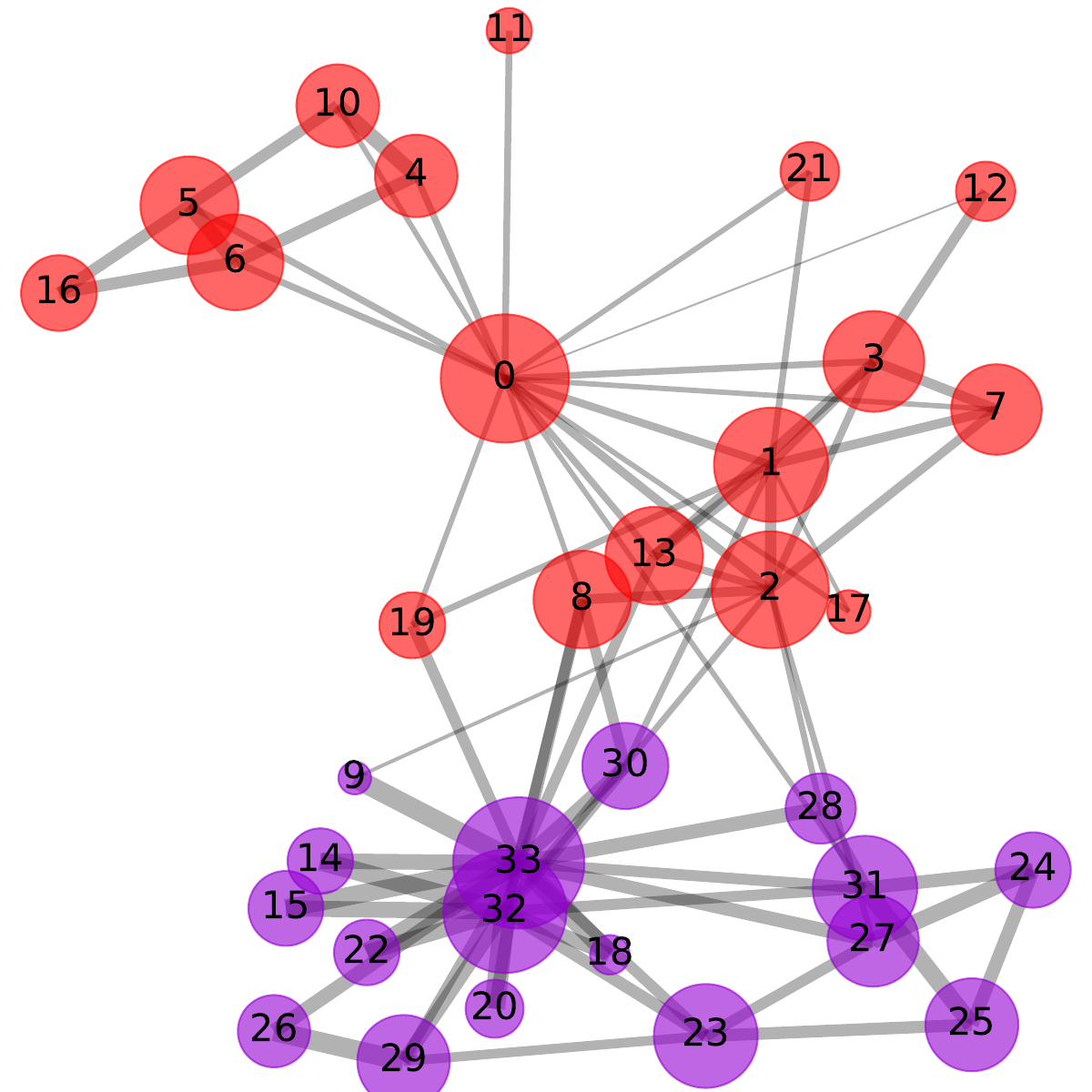} 
        \caption{\shortstack{Original \\ $[0.52, 0.48]$}}
        \label{fig:graph_original}
    \end{subfigure}
    \hfill
    \begin{subfigure}[b]{0.164\textwidth}
        \centering
        \includegraphics[width=\textwidth]{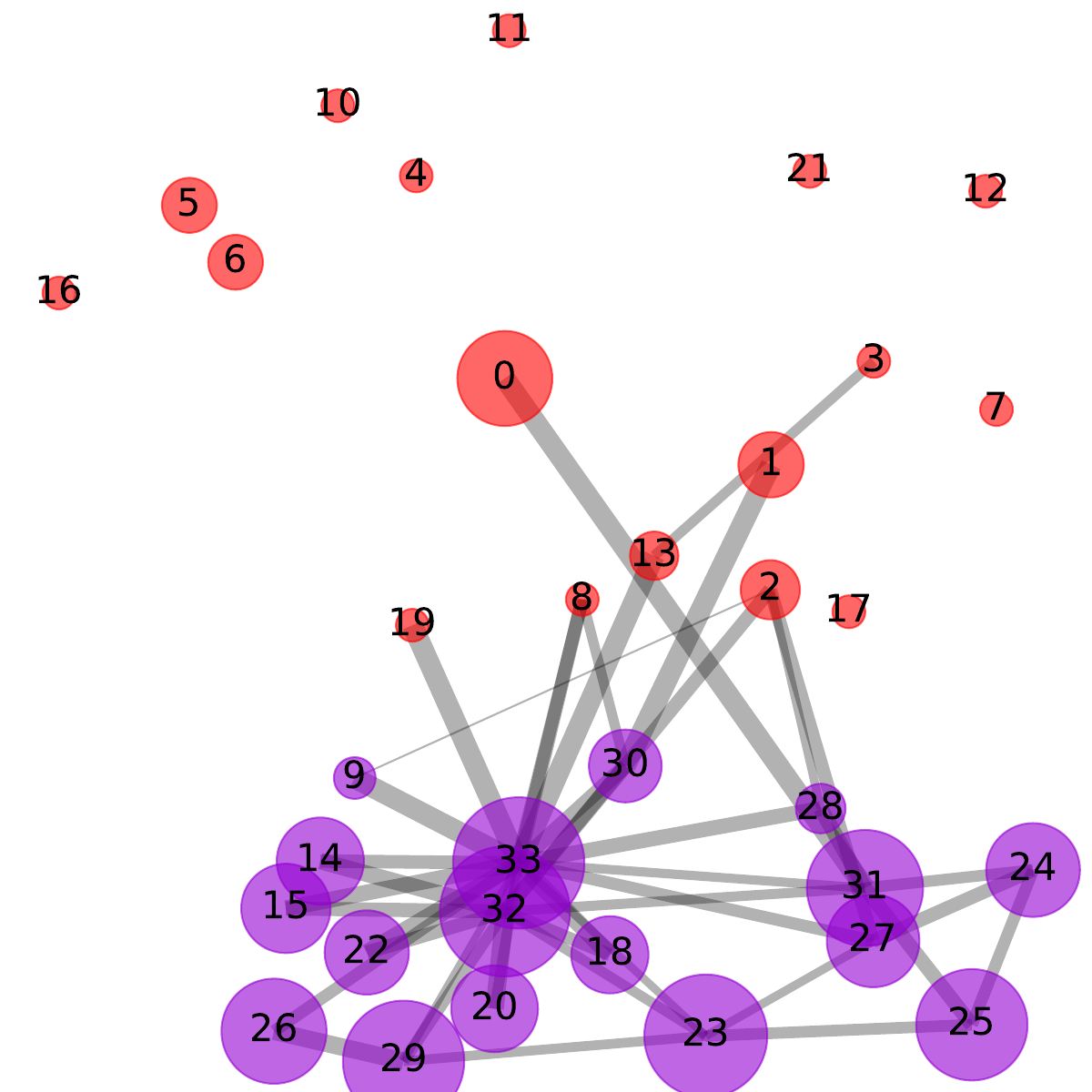}
        \caption{\shortstack{\fairgd \\ $[0.12,0.88]$}}
        \label{fig:graph_fairgd}
    \end{subfigure}
    \hfill
    \begin{subfigure}[b]{0.164\textwidth}
        \centering
        \includegraphics[width=\textwidth]{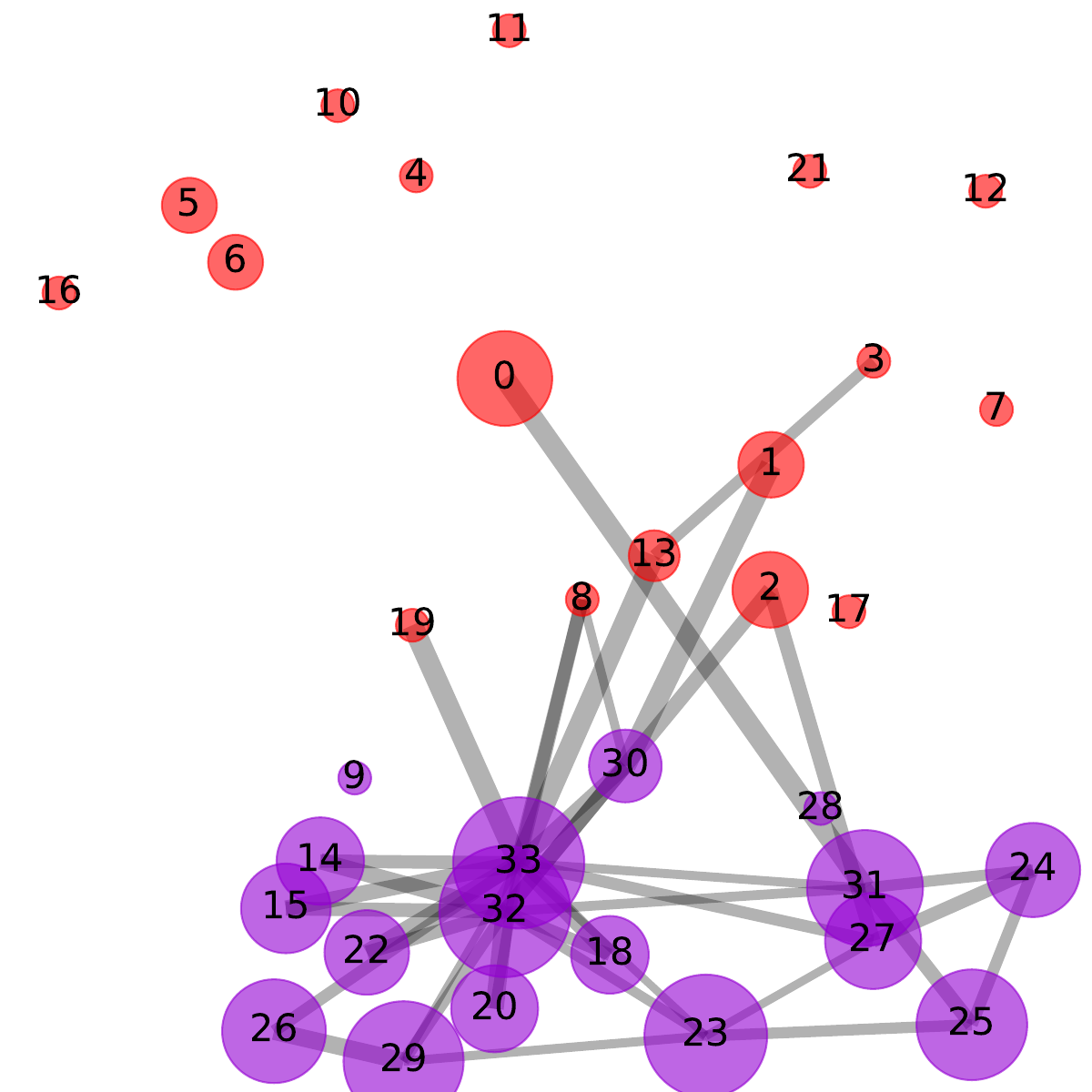}
        \caption{\shortstack{\crossgd \\ $[0.13,0.87]$}}
        \label{fig:graph_adapted}
    \end{subfigure}
    \hfill
    \begin{subfigure}[b]{0.164\textwidth}
        \centering
        \includegraphics[width=\textwidth]{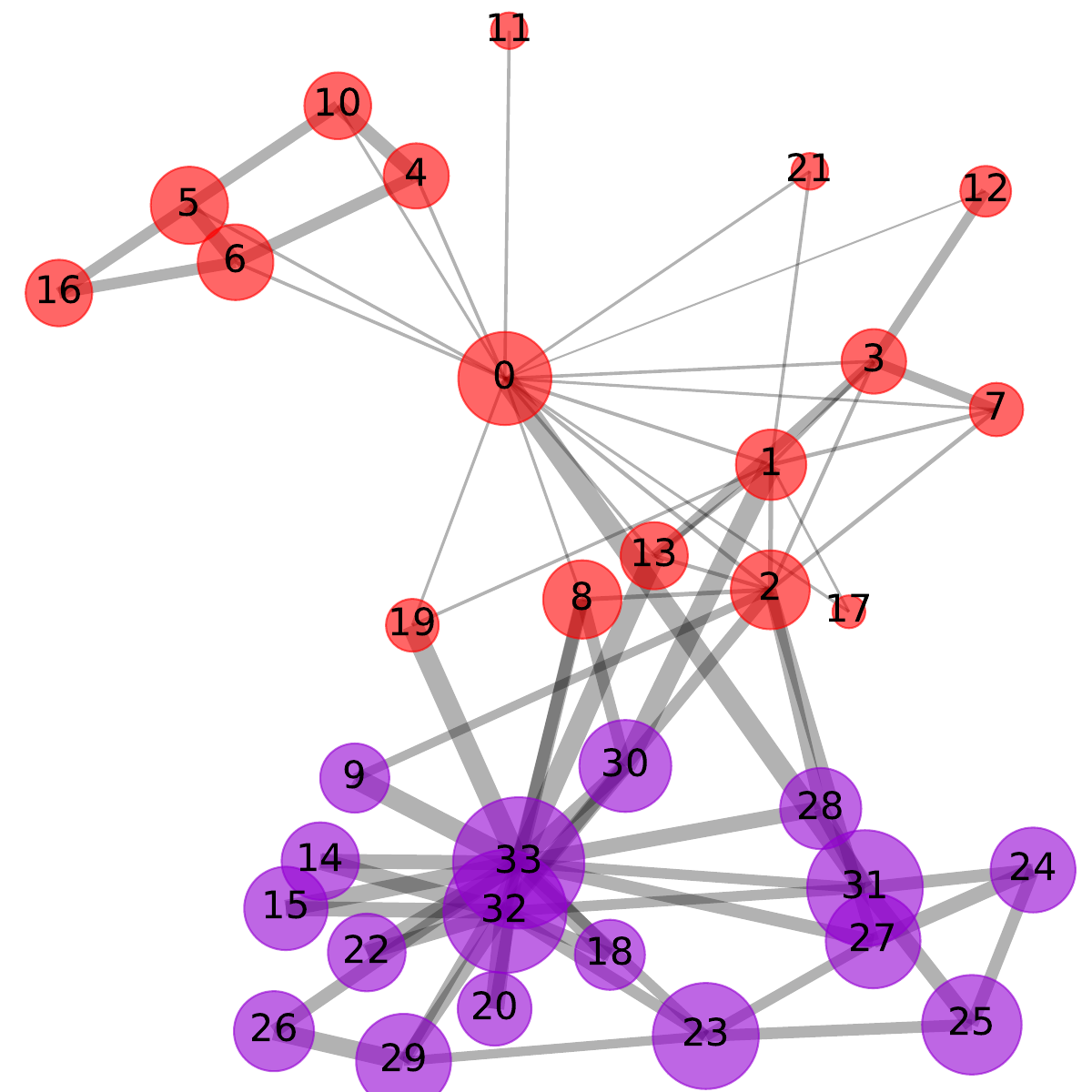}
        \caption{\shortstack{\fairwalk \\ $[0.22,0.78]$}}
        \label{fig:graph_fw}
    \end{subfigure}
    \hfill
    \begin{subfigure}[b]{0.164\textwidth}
        \centering
        \includegraphics[width=\textwidth]{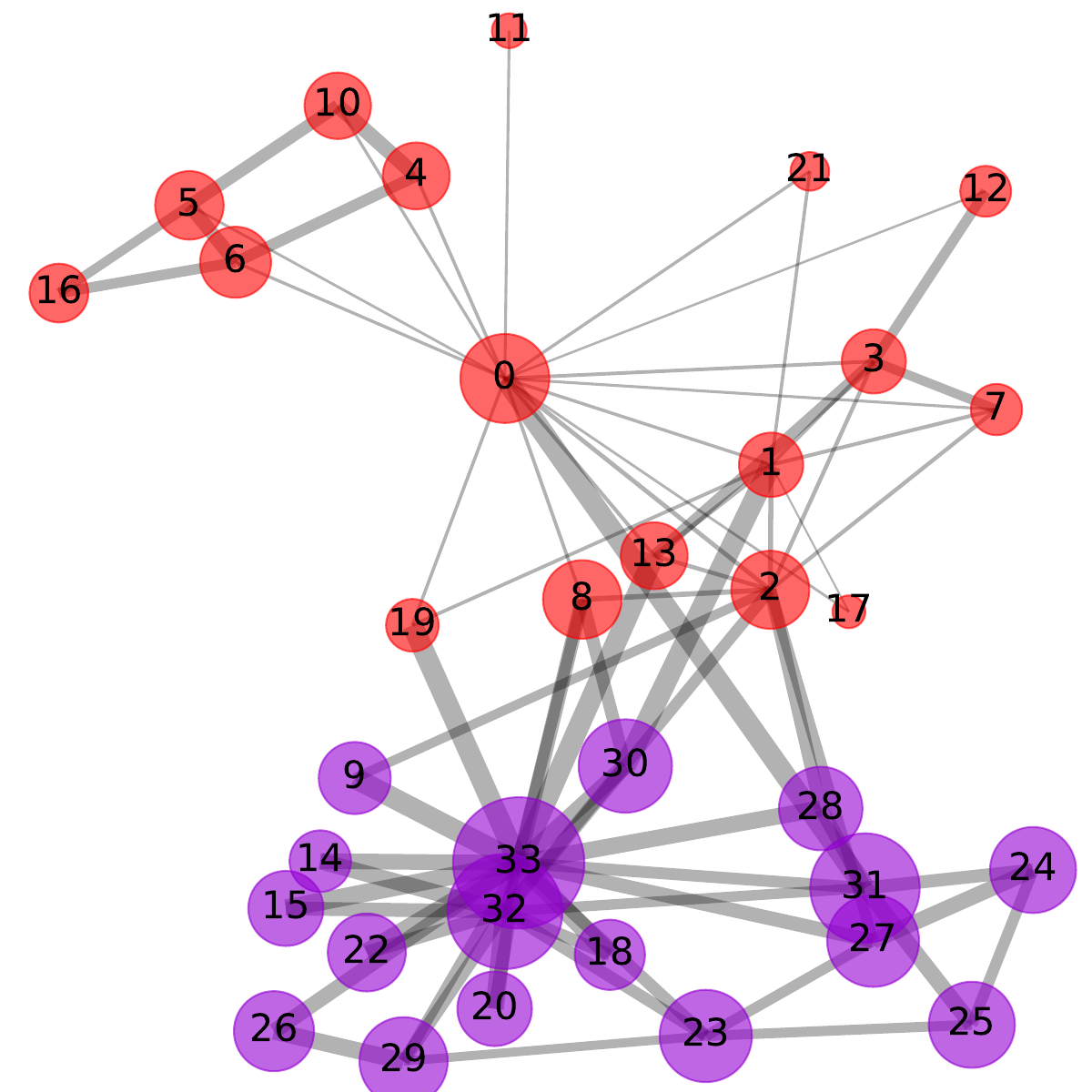}
        \caption{\shortstack{\crosswalk \\ $[0.22,0.78]$}}
        \label{fig:graph_cw}
    \end{subfigure}

    \vspace{2mm} 

    \begin{subfigure}[b]{0.163\textwidth}
        \centering
        \includegraphics[width=\textwidth]{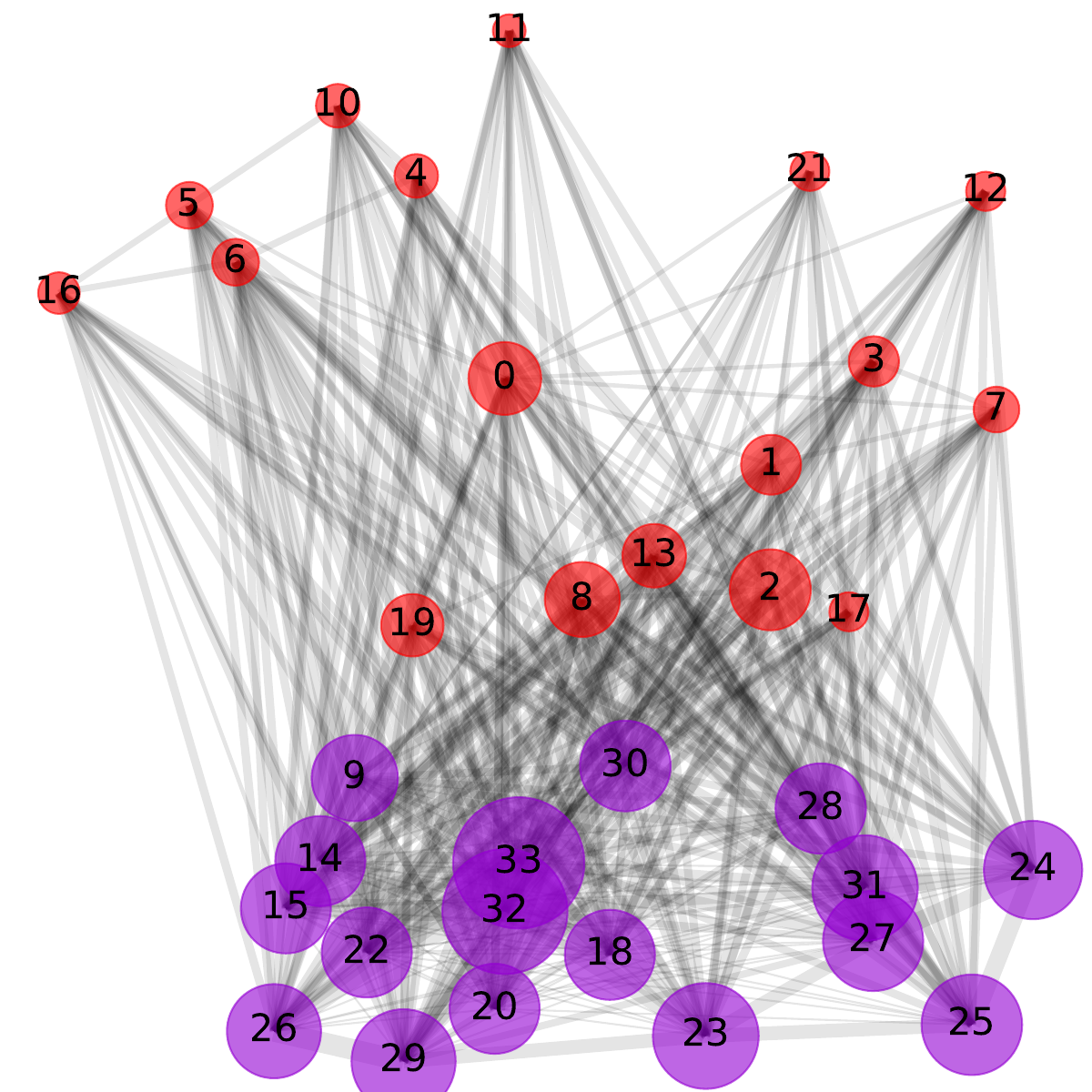} 
        \caption{\shortstack{\lfpru \\ $[0.16,0.84]$}}
        \label{fig:fairone}
    \end{subfigure}
    \hfill
    \begin{subfigure}[b]{0.163\textwidth}
        \centering
        \includegraphics[width=\textwidth]{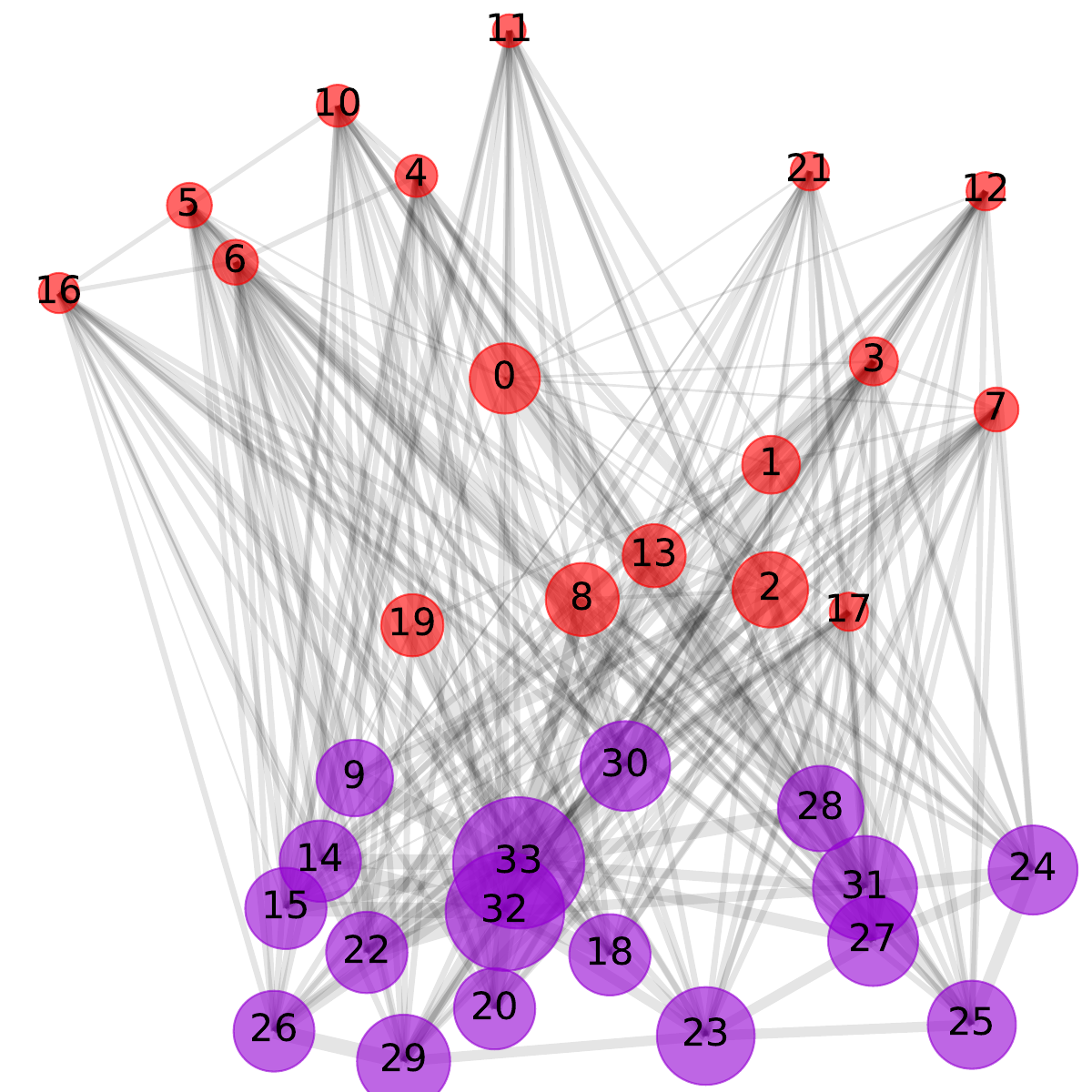}
        \caption{\shortstack{\lfprn \\ $[0.16, 0.84]$}}
        \label{fig:graph_fairtwo}
    \end{subfigure}
    \hfill
    \begin{subfigure}[b]{0.163\textwidth}
        \centering
        \includegraphics[width=\textwidth]{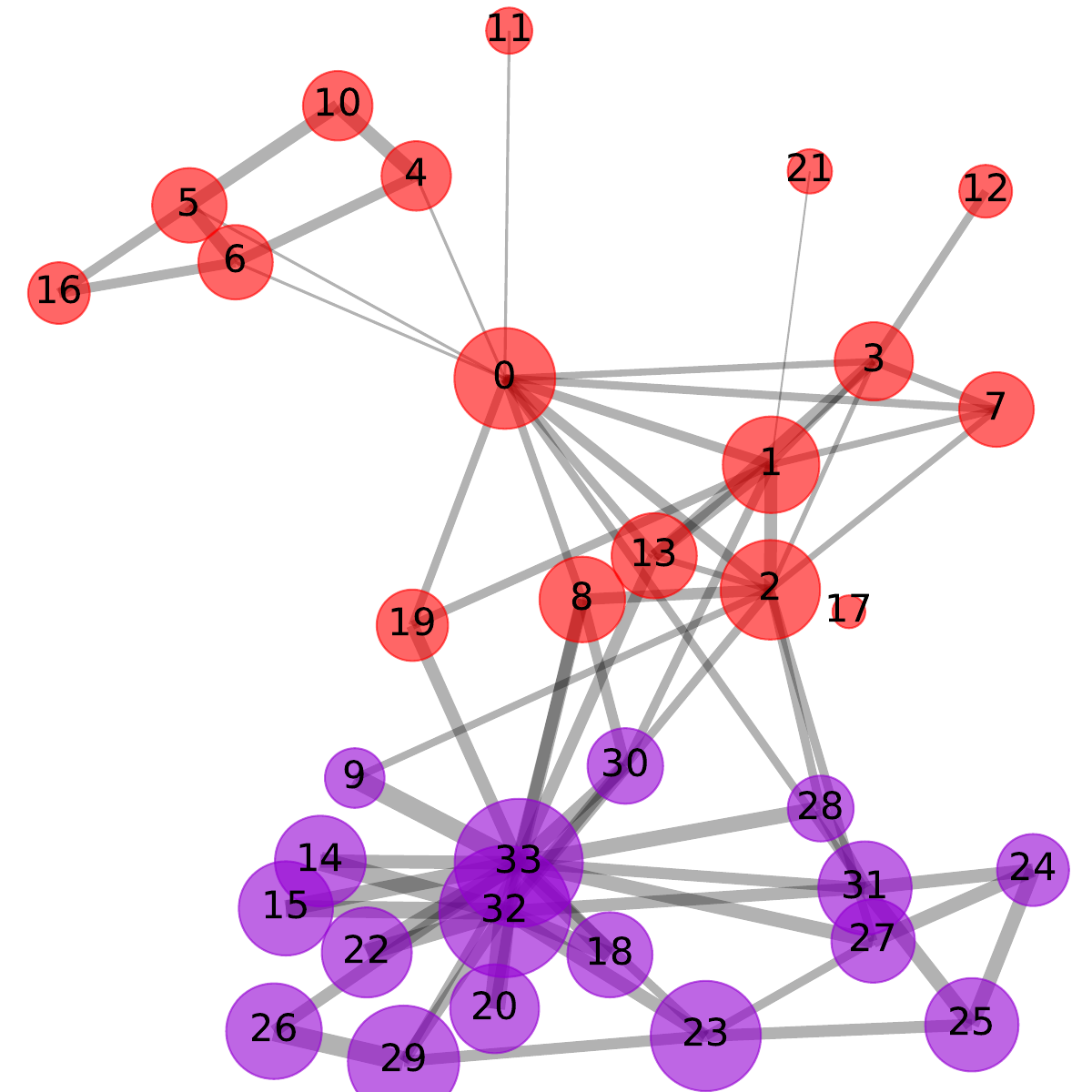} 
        \caption{\shortstack{\fairgd($0.1$,$0.05$) \\ $[0.30,0.70]$}}
        \label{fig:fairone_var}
    \end{subfigure}
    \hfill
    \begin{subfigure}[b]{0.163\textwidth}
        \centering
        \includegraphics[width=\textwidth]{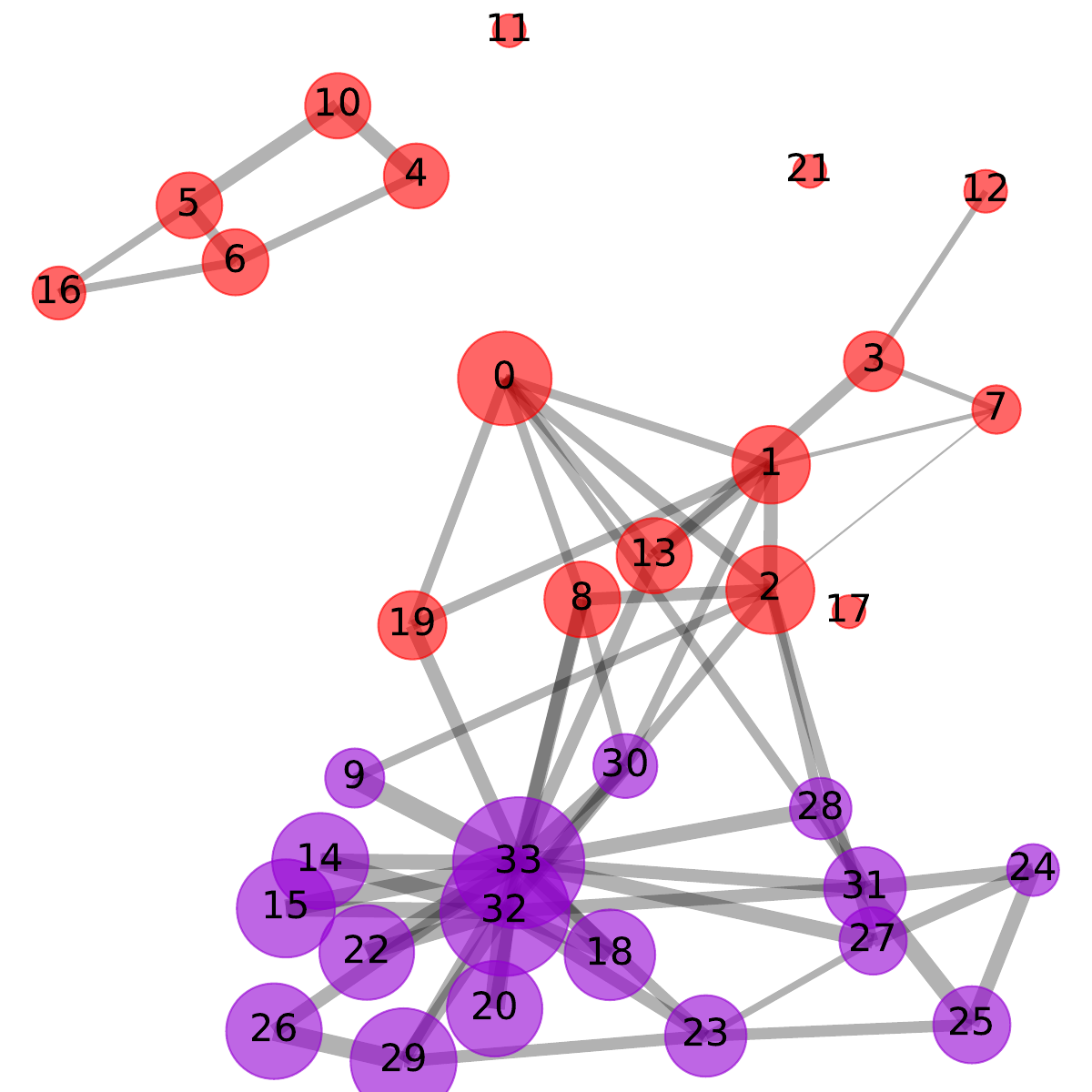}
        \caption{\shortstack{\fairgd($0.1$,$0.1$) \\ $[0.22, 0.78]$}}
        \label{fig:graph_fairtwo_var}
    \end{subfigure}
    \hfill
    \begin{subfigure}[b]{0.163\textwidth}
        \centering
        \includegraphics[width=\textwidth]{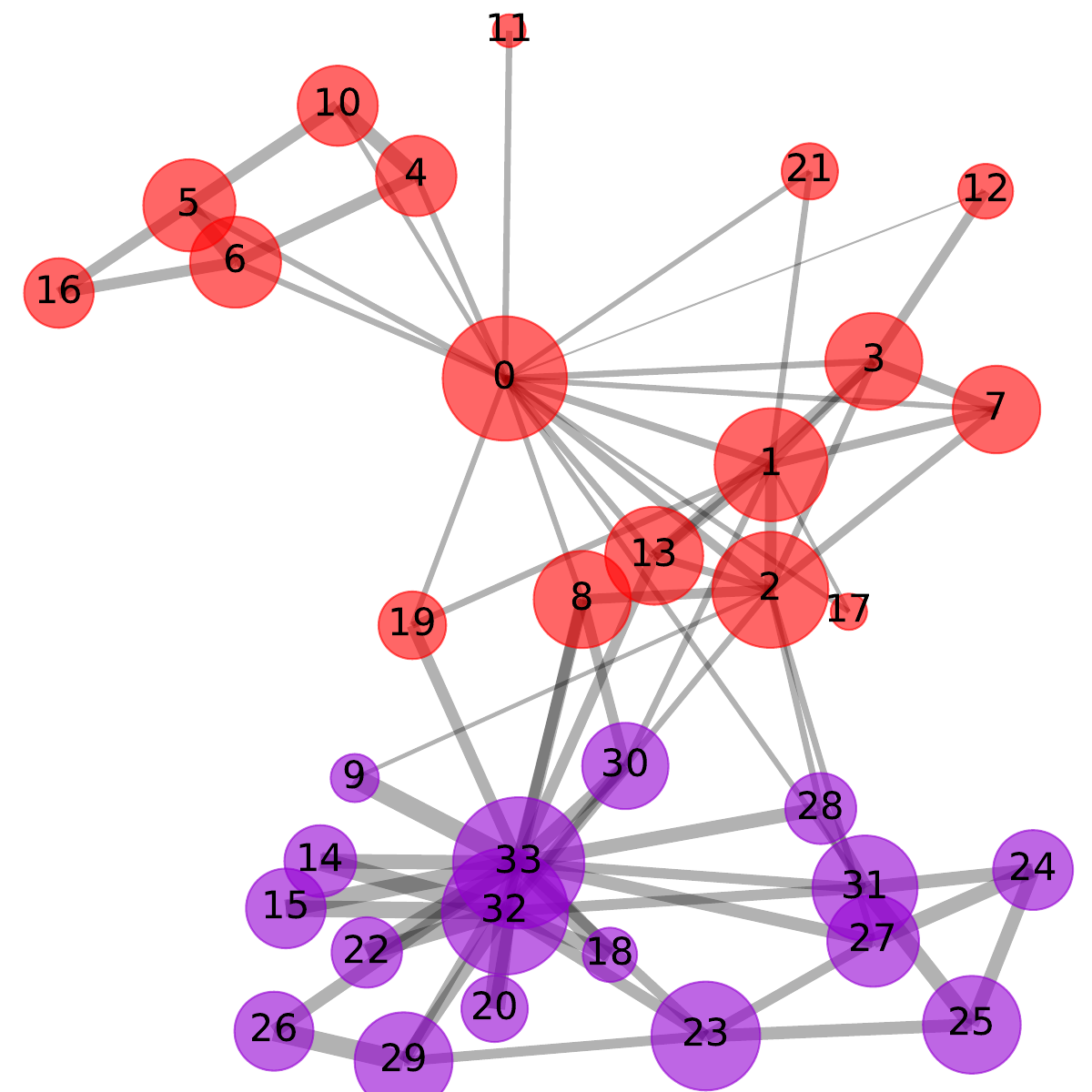}
        \caption{\shortstack{\crossgd($0.1$,$0.05$) \\ $[0.48,0.52]$}}
        \label{fig:graph_adaptedone}
    \end{subfigure}
    \hfill
    \begin{subfigure}[b]{0.163\textwidth}
        \centering
        \includegraphics[width=\textwidth]{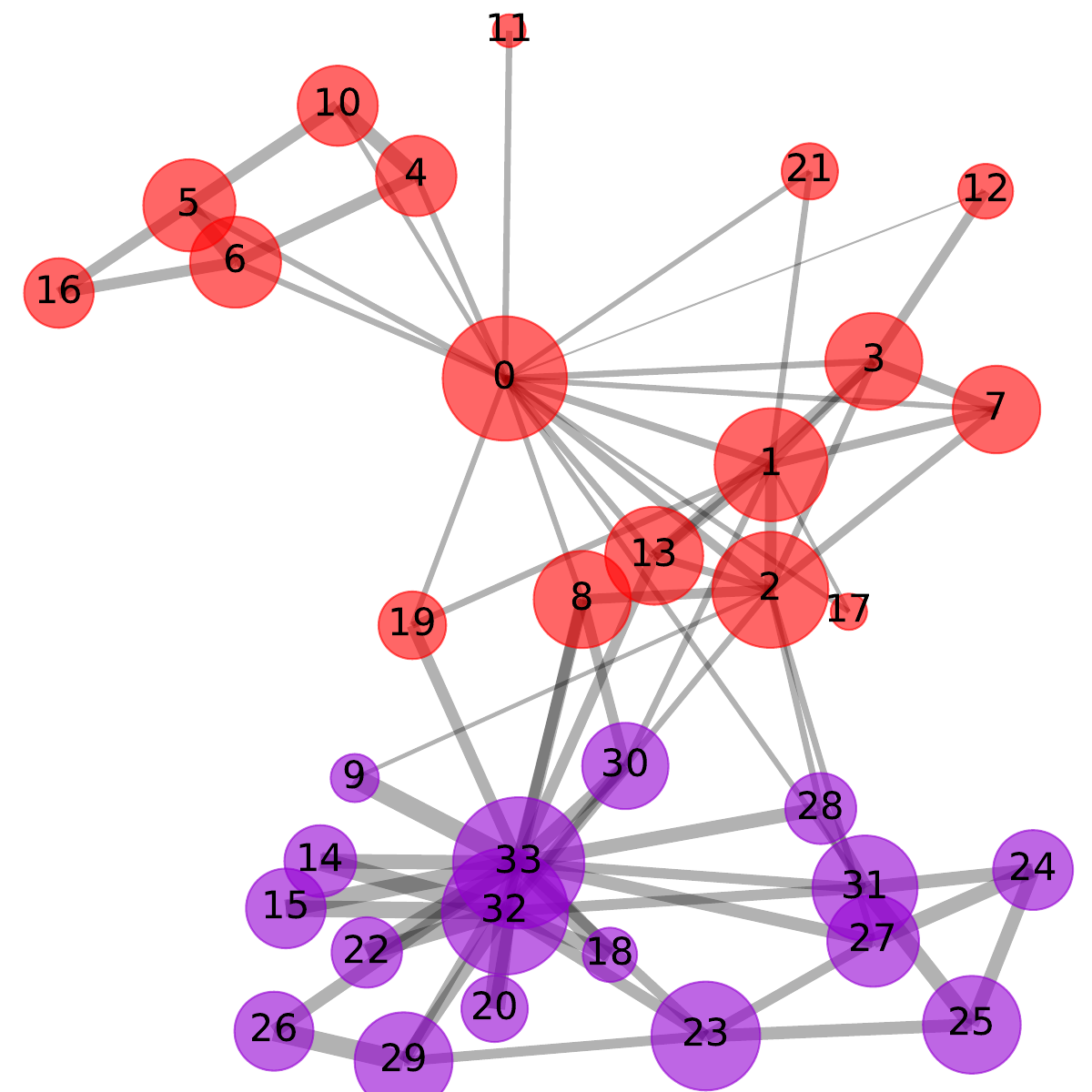}
        \caption{\shortstack{\crossgd($0.1$,$0.1$) \\ $[0.48,0.52]$}}
        \label{fig:graph_adapttwo}
    \end{subfigure}

    \caption{
    Graph structure of Karate before and after edge reweighting by each method. The target group-wise fairness vector is $[0.1,0.9]$. Below each method name, the group-wise \pagerank scores calculated using a uniform restart vector are displayed. The two values correspond to the red and purple groups, respectively.
    }
    \label{fig:graph_visualization}
\end{figure*}


\subsection{Fairness loss and weight change \texorpdfstring{$\Delta_\transmatrix$}{Delta $P$}}

We evaluate the fairness loss $\loss(\newtransmatrix, \gamma, \vectv)$
on \fairgd, {\fairgd}($0.1$, $0.1$), {\fairgd}($0.5$, $0.1$), and all baselines across the five datasets; 
the results are shown in \Cref{fig:loss_all_dataset}. 
We also evaluate the changes, denoted as $\Delta_\transmatrix$, to the original transition matrix; 
the results are shown in \Cref{fig:change_all_dataset}.
The results demonstrate that \fairgd and its restricted variants generally achieve better or comparable fairness loss compared to the baselines, while maintaining significantly smaller changes in the transition matrix weights.

On the two-group datasets, \lfpru and \lfprn achieve near-perfect fairness by modifying the graph so that each node fairly distributes its \pagerank score to its neighbors according to the target fairness vector. On the \blogs dataset, \fairgd and its two restricted variants exhibit very similar behavior, achieving 8 out of the 10 best fairness outcomes while also minimizing weight changes. For the \twitter dataset, \fairgd and its variants outperform \fairwalk and \crosswalk in terms of fairness loss, whereas \fairwalk performs comparably with our proposed methods in weight change, surpassing all other baselines. On the \mind dataset, \fairgd and all baselines achieve almost perfect fairness, but \fairgd consistently outperforms them in weight change. However, for \(\phi > 0.5\), the constrained variants \fairgdrone and \fairgdrtwo fail to maintain high fairness but consistently achieve smaller weight changes.

For multi-group datasets, \fairgd and its two variants perform best on both fairness and weight change on the \slashdot dataset. On the \book dataset, \fairgd attains the highest fairness score, followed by \fairgdrtwo and \fairgdrone; the weight changes surpass those of \fairwalk and \crosswalk for some $\phi$ values to achieve better fairness outcomes.





Overall, \fairgd delivers superior results in terms of fairness loss and weight change when \grouppr is close to the original PageRank score, $\grouppr_1$ (the first dimension of $\groupDist_o$ as reported in \Cref{table:datasets}). As \grouppr diverges from $\grouppr_1$, both the fairness loss and weight change increase. The two restricted variants, \fairgdrone and \fairgdrtwo, exemplify the trade-off between fairness loss and weight change by achieving smaller weight changes at the cost of higher fairness loss compared to \fairgd.
This refinement highlights the balance between fairness and weight change, 
emphasizing the flexibility of \fairgd in adapting to various fairness~goals.

\subsection{Group-adapted fairness loss and \texorpdfstring{$\Delta_\transmatrix$}{Delta $P$}}

We evaluate \crossgd and its restricted variants, \crossgdrone and \crossgdrtwo, on the group-adapted fairness loss, as defined in \Cref{equation:cgfl}, and analyze the corresponding changes in the transition matrix weights. 
The results 
are presented in \Cref{fig:loss_all_dataset_adapted} for the group-adapted fairness loss 
and in \Cref{fig:change_all_dataset_adapted} for the transition matrix weight changes.
The results show that the trade-off between fairness loss and weight change is further exemplified and shown for \crossgd and its variants.



Across the \book, \blogs, and \mind datasets, \crossgd consistently achieves the most favorable fairness outcomes compared to competing baselines. In terms of weight change, \crossgd attains optimal performance on the \book dataset and 9 out of 10 optimal outcomes on the \mind dataset. Although \crossgd underperforms \lfpru and \lfprn in fairness loss on the \twitter dataset, it still matches or surpasses the other two baseline methods, while delivering substantially better weight change results. On the \slashdot dataset, \crossgd outperforms all competing baselines in fairness outcome, and achieves the best weight change metrics for 4 out of 10 values of $\phi$. Notably, on the \blogs and the \slashdot dataset, \crossgd prioritizes improvements in fairness over weight change performance.

For all datasets, the two variants of \crossgd exhibit comparable or slightly higher fairness loss relative to \crossgd, while consistently maintaining lower or comparable weight change.

\subsection{Ranking coefficient evaluation}
We report the average Spearman’s rank correlation coefficient $\overline{\rho}$ for each method across multiple datasets. For each group, we compute the Spearman’s coefficient between the rankings of old and new PageRank values restricted to that group. These per-group coefficients are then aggregated using a weighted average, where each group’s weight corresponds to its proportion of the total items. The results for the \mind dataset are shown in \Cref{table:ranking_mind}, while results for other datasets are provided in Appendix \ref{appendix:more-experiments}.
In the tables, we depict with boldface the largest value and with underline the second largest.

Across all $\phi$ values, \fairgd consistently attains the highest $\overline{\rho}$, followed by \fairgdrtwo and \fairgdrone.
These results indicate that \fairgd exhibits very good preservation of the relative PageRank score rankings within individual groups. These results suggest that, in practical applications, when adjusting the content displayed to users based on items’ PageRank scores to enhance fairness, fewer changes are required within
each group. Instead, greater focus should be placed on adjusting
the frequency or weighting of content across different groups.
\subsection{Use case}

We visualize the graph structure of a small real-world dataset, Karate \citep{zachary1977information, nr}, before and after edge reweighting in \Cref{fig:graph_visualization}. The Karate dataset is a classic social network from Wayne Zachary’s 1977 study of friendships among 34 members of a karate club, which eventually split into two factions after a disagreement. The dataset contains 34 nodes, with 17 nodes belonging to the ``Mr. Hi" club and the remaining 17 nodes belonging to the ``Officer" club.

In the visualization, vertex labels indicate node indices, node colors correspond to group membership, node sizes are proportional to their \pagerank scores, and edge thickness reflects transition weights.

The original group-wise \pagerank score for the Karate dataset is $[0.52, 0.48]$, while the fairness goal is set to $[0.1, 0.9]$. From \Cref{fig:graph_visualization}, we observe that \fairgd achieves the closest group-wise fairness scores, followed by \crossgd, \lfpru, and \lfprn. The unrestricted versions of \fairgd and \crossgd tend to remove a large proportion of intra-group edges within the red group, whereas their restricted counterparts delete fewer edges due to the modification bounds, at the cost of achieving a lower fairness improvement. \fairwalk and \crosswalk remove no edges, resulting in graph structures very similar to the original. Finally, \lfpru modifies the graph to be fully connected, achieving the same fairness score as \lfprn, which adds fewer non-existing edges.

\section{Conclusion}

We presented a novel approach for incorporating group
fairness into the PageRank algorithm by re\-weighting the entries of the underlying transition matrix. 
We focus on matrix re\-weighting interventions, 
as we argue that those are more actionable strategies,
and further, reflect the network structure, rather than the hyper-parameters of \pagerank.
Our goal is to minimize fairness loss (with respect to a target group-fairness distribution)
while also minimizing the changes to the transition matrix. 
We achieve our objective by foregoing the idea of ensuring fairness locally, at every vertex, 
and instead aim to ensure fairness at a global level.
Technically, our method is an efficient projected gradient-descent algorithm for
finding locally-optimal edge weights.
We presented a comprehensive empirical evaluation, 
demonstrating that our method offers significant advantages compared to strong~baselines.

\begin{acks}
We sincerely thank Sebastian Dalleiger for providing constructive feedback.
This research is supported by 
the ERC Advanced Grant REBOUND (834862),
the Swedish Research Council project ExCLUS (2024-05603),
and the Wallenberg AI, Autonomous Systems and Software Program (WASP) funded by the Knut and Alice Wallenberg Foundation.
\end{acks}

\section{Ethical considerations}

This work focuses on reweighting the transition matrix of a graph to achieve target group-wise PageRank scores, with the goal of improving fairness across groups. While the proposed approach aims to improve fairness and mitigate bias in ranking systems, interventions that alter item rankings could be misused to deliberately amplify or suppress the visibility of certain content, potentially leading to manipulation or censorship. Furthermore, if applied to sensitive domains, such interventions could unintentionally disadvantage specific users or groups, even when fairness metrics improve. Our experiments use only publicly available datasets with no personal user or item-identifying information, reducing privacy risks. To mitigate potential misuse, we recommend deploying such methods with transparent governance, independent auditing of fairness outcomes, and clear documentation of intervention goals and criteria.

\balance

\bibliographystyle{ACM-Reference-Format}
\bibliography{refs.bib}

\clearpage
\appendix
\section{Proof of Proposition~\ref{proposition:non-convexity}}
\label{appendix:omitted-proposition-non-convexity}

To demonstrate the non-convexity, it suffices to provide a single counterexample. Consider a fully connected graph $\graph$ with three vertices: $a$, $b$, and $c$. Let the labels be $\labelof_a =\mathsf{blue}$, and $\labelof_b = \labelof_c = \mathsf{red}$. We also let 
\[ 
\newtransmatrix_1 =  \begin{pmatrix}
0 & 1 & 0 \\
1 & 0 & 0 \\
0 & 1 & 0 \\
\end{pmatrix} 
~\mbox{ and }~ 
\newtransmatrix_2 =  \begin{pmatrix}
0 & 0 & 1 \\
0 & 0 & 1 \\
1 & 0 & 0 \\
\end{pmatrix}\] 
be two row-stochastic transition matrices, and 
\[
\newtransmatrix_3 = \frac{1}{2}\newtransmatrix_1 + \frac{1}{2}\newtransmatrix_2 = \begin{pmatrix}
0 & 1/2 & 1/2 \\
1/2 & 0 & 1/2 \\
1/2 & 1/2 & 0 \\
\end{pmatrix}.
\]

Given restart vector $\vectv = (1/3, 1/3, 1/3)$ and restart probability $\gamma$, 
we can calculate the \pagerank vectors $\prvector_1$, $\prvector_2$, and $\prvector_3$, 
for $\newtransmatrix_1$, $\newtransmatrix_2$, and $\newtransmatrix_3$, 
being 
$\prvector_1 = \left(\frac{\gamma^2-3\gamma+3}{3(2-\gamma)}, \frac{3-2\gamma}{3(2-\gamma)}, \frac{\gamma}{3}\right)$, 
$\prvector_2=\left(\frac{\gamma^2-3\gamma+3}{3(2-\gamma)},  \frac{\gamma}{3}, \frac{3-2\gamma}{3(2-\gamma)},\right)$, and 
$\prvector_3=\left( \frac{1}{3},\frac{1}{3},\frac{1}{3}\right)$, respectively. 

Now, consider the group-wise target \pagerank score, defined as $\grouppr_1 = \frac{\gamma^2-3\gamma+3}{3(2-\gamma)}$ and $\grouppr_2 = 1-\grouppr_1$. It holds that $\frac{1}{2}\loss(\newtransmatrix_1, \gamma, \vectv) + \frac{1}{2}\loss(\newtransmatrix_2, \gamma, \vectv) = 0 $. One can easily find a $\gamma$ such that that $\loss(\frac{1}{2}\newtransmatrix_1 + \frac{1}{2}\newtransmatrix_2, \gamma, \vectv) = \loss(\newtransmatrix_3, \gamma, \vectv) >0$, which contradicts the definition of convexity. This completes the proof.

\section{Proof of Proposition~\ref{theorem:gradient-fairness-loss}}
\label{appendix:omitted-proposition-proof}

\phipagerankgradient*
Recalling that $\prvector = \gamma \matrixU^{-1} \vectv$,
we have the following:
\begin{equation*}
\begin{aligned}
\label{eq:gradient_fairness_loss}
    \frac{\partial}{\partial \transmatrix} \loss(\transmatrix, \gamma, \vectv) 
        & = \frac{\partial}{\partial \transmatrix} \frac{1}{K}
                \sum_{k=1}^{\nogroups}  \left(\indvect{\agroup}^{\mathsf{T}} \prvector - \grouppr_\agroup\right)^2 \\
        & = \frac{2}{K} \sum_{k=1}^{\nogroups} \left(\indvect{\agroup}^{\mathsf{T}} \prvector - \grouppr_\agroup\right) 
                \frac{\partial}{\partial \transmatrix} \left(\indvect{\agroup}^{\mathsf{T}} \prvector - \grouppr_\agroup\right) \\
        & = \frac{2}{\nogroups} \sum_{k=1}^{\nogroups} \left(\indvect{\agroup}^{\mathsf{T}} \prvector - \grouppr_\agroup\right) 
                \frac{\partial}{\partial \transmatrix} \left(\indvect{\agroup}^{\mathsf{T}} \prvector\right)  \\ 
        & = \frac{2 \gamma}{K} \sum_{k=1}^{\nogroups} \left(\indvect{\agroup}^{\mathsf{T}} \prvector - \grouppr_\agroup\right) 
                \frac{\partial}{\partial \transmatrix} \left(\indvect{\agroup}^{\mathsf{T}} \matrixU^{-1} \vectv\right). 
\end{aligned}
\end{equation*}
By the chain rule of matrix calculus \citep[Equation~(137)]{petersen_matrix_2008}, 
we obtain:
\begin{equation}
\label{eq:chain_rule}
\begin{aligned}
\frac{\partial}{\partial \transmatrix[i, j]} \left(\indvect{\agroup}^{\mathsf{T}} \matrixU^{-1} \vectv\right) 
    & = \mathrm{Tr} 
        \left[ 
            \left( \frac{\partial (\indvect{\agroup}^{\mathsf{T}} \matrixU^{-1} \vectv)}{\partial \matrixU} \right)^{\mathsf{T}} 
            \frac{\partial \matrixU}{\partial \transmatrix[i, j]} 
        \right]
\end{aligned}
\end{equation}
By considering the derivative of inverse~\cite[Equation~(61)]{petersen_matrix_2008}~we~have:
\begin{equation*}
    \frac{\partial (\indvect{\agroup}^{\mathsf{T}} \matrixU^{-1} \vectv)}{\partial \matrixU} 
        = -\matrixU^{-T} \indvect{\agroup} \vectv^{\mathsf{T}} \matrixU^{-T}.
\end{equation*}
The gradient of $\matrixU$ with respect to $\transmatrix[i, j]$ is:
\begin{equation*}
    \frac{\partial \matrixU}{\partial \transmatrix[i, j]} 
        = - (1 - \gamma) \frac{\partial \transmatrix^{\mathsf{T}}}{\partial \transmatrix[i, j]} = - (1 - \gamma) \Ematrix_{ji},
\end{equation*}
where $\Ematrix_{ij}$ is the matrix whose $[i,j]$ entry is 1 and all other entries are 0.
From Equation~\eqref{eq:chain_rule}, we get:
\begin{equation*}
\label{eq:gradient_onek_Uinv_v}
\begin{aligned}
\frac{\partial}{\partial \transmatrix[i, j]} \left(\indvect{\agroup}^{\mathsf{T}} \matrixU^{-1} \vectv\right) 
    & = (1 - \gamma) \mathrm{Tr} 
            \left[ \left( \matr{U}^{-T} \indvect{\agroup} \vectv^{\mathsf{T}} \matrixU^{-T} \right)^{\mathsf{T}} \Ematrix_{ji} \right] \\ 
    & = (1 - \gamma) \left( \matrixU^{-1} \vectv \indvect{\agroup}^{\mathsf{T}} \matrixU^{-1} \right)[i, j] \\
    & = \frac{1 - \gamma}{\gamma} \left( \prvector \indvect{\agroup}^{\mathsf{T}} \matrixU^{-1} \right)[i, j].
\end{aligned}
\end{equation*}

Substituting~\eqref{eq:gradient_onek_Uinv_v} into Equation~\eqref{eq:gradient_fairness_loss} 
leads to 
\begin{equation}
\begin{aligned}
\label{eq:gradient_fairness_loss_final}
    \frac{\partial}{\partial \transmatrix} \loss(\transmatrix, \gamma, \vectv) 
         = \frac{2 (1-\gamma)}{K} \sum_{k=1}^{\nogroups} \left(\indvect{\agroup}^{\mathsf{T}} \prvector - \grouppr_\agroup\right) \left( \prvector \indvect{\agroup}^{\mathsf{T}} \matrixU^{-1} \right)           
\end{aligned}
\end{equation}


\section{Proof of Proposition~\ref{theorem:adapted-gradient-fairness-loss}}
\label{appendix:omitted-proposition-adapted-gradients}

\gaphipagerankgradient*

Similarly to the proof in Appendix \ref{appendix:omitted-proposition-proof}, for the group-adapted fairness loss, we can compute its gradient over $\transmatrix$ as follows:
\begin{equation}
\begin{aligned}
\label{eq:gradient_fairness_loss_group_adapted}
    \frac{\partial}{\partial \transmatrix} \groupAdapLoss(\transmatrix, \gamma) 
        & = \frac{\partial}{\partial \transmatrix} \frac{1}{\nogroups^2}
    \sum_{k=1}^{\nogroups}\sum_{l=1}^{\nogroups} \left(\indvect{\agroup}^{\mathsf{T}} \prvector_{\ell} - \grouppr_\agroup\right)^2 \\
        & = \frac{2}{\nogroups^2} \sum_{k=1}^{\nogroups} \sum_{l=1}^{\nogroups}\left(\indvect{\agroup}^{\mathsf{T}} \prvector_{\ell} - \grouppr_\agroup\right) 
                \frac{\partial}{\partial \transmatrix} \left(\indvect{\agroup}^{\mathsf{T}} \prvector_{\ell} - \grouppr_\agroup\right) \\
        & = \frac{2}{\nogroups^2} \sum_{k=1}^{\nogroups} \sum_{l=1}^{\nogroups}\left(\indvect{\agroup}^{\mathsf{T}} \prvector_{\ell} - \grouppr_\agroup\right) 
                \frac{\partial}{\partial \transmatrix} \left(\indvect{\agroup}^{\mathsf{T}} \prvector_{\ell} \right)  \\ 
        & = \frac{2 \gamma}{\nogroups^2} \sum_{k=1}^{\nogroups} \sum_{l=1}^{\nogroups}\left(\indvect{\agroup}^{\mathsf{T}} \prvector_{\ell} - \grouppr_\agroup\right) 
                \frac{\partial}{\partial \transmatrix} \left(\indvect{\agroup}^{\mathsf{T}} \matrixU^{-1} \vectv_{\ell} \right). 
\end{aligned}
\end{equation}

According to \Cref{eq:gradient_onek_Uinv_v},

\begin{equation}
\label{eq:gradient_onek_Uinv_v_l}
\begin{aligned}
\frac{\partial}{\partial \transmatrix[i, j]} \left(\indvect{\agroup}^{\mathsf{T}} \matrixU^{-1} \vectv_{\ell} \right) 
    & = (1 - \gamma) \left( \matrixU^{-1} \vectv_{\ell} \indvect{\agroup}^{\mathsf{T}} \matrixU^{-1} \right)[i, j] \\
    & = \frac{1 - \gamma}{\gamma} \left( \prvector_{\ell}\indvect{\agroup}^{\mathsf{T}} \matrixU^{-1} \right)[i, j].
\end{aligned}
\end{equation}
Substituting \Cref{eq:gradient_onek_Uinv_v_l} to \Cref{eq:gradient_fairness_loss_group_adapted}, we get

\begin{equation}
\begin{aligned}
\label{eq:gradient_cross_fairness_loss}
    \frac{\partial}{\partial \transmatrix} \groupAdapLoss(\transmatrix, \gamma) 
         = \frac{2 (1-\gamma)}{K^2} \sum_{k=1}^{\nogroups} 
         \sum_{\ell=1}^{\nogroups} \left(\indvect{\agroup}^{\mathsf{T}} \prvector_{\ell} - \grouppr_\agroup\right) \left( \prvector_{\ell} \indvect{\agroup}^{\mathsf{T}} \matrixU^{-1} \right)         
\end{aligned}
\end{equation}

\section{Approximating $\vecty_k$}
\label{appendix:Neumann series}

In this section, we prove that for approximating $\vecty_t$, truncating the Neumann series to its first $50$ terms yields a relative error below $0.0003$.

According to the Neumann series formula, we can calculate the matrix inverse of a matrix $\unitary - \textbf{X}$ as: 
$
(\textbf{I} - \textbf{X})^{-1} = \sum_{t=0}^\infty \textbf{X}^{\mathsf{T}}.
$ 
Let $\textbf{S}_t$ denote the partial sum of the Neumann series up to the $t$-th term, that is,
\begin{equation*}
    \textbf{S}_t = \sum_{i=0}^t \textbf{X}^{\mathsf{T}} = \textbf{I} + \textbf{X} + \textbf{X}^2 + \dots + \textbf{X}^k.
\end{equation*}
We denote the truncation error as
$
\textbf{E}_t = (\textbf{I} - \textbf{X})^{-1} - \textbf{S}_t
$, then we can express $\textbf{E}_t$ as the sum of the remaining terms:
\begin{equation*}
    \textbf{E}_t = \sum_{i=t+1}^\infty \textbf{X}^{\mathsf{T}}
\end{equation*}
By factoring out $\textbf{X}^{t+1}$ from the series we get:
\begin{equation*}
    \textbf{E}_t = \textbf{X}^{t+1} \sum_{i=0}^\infty \textbf{X}^{\mathsf{T}}.
\end{equation*}
Using the formula for the Neumann series, the infinite sum $\sum_{t=0}^\infty \textbf{X}^{\mathsf{T}}$ is equal to $(\textbf{I} - \textbf{X})^{-1}$. Thus:
\begin{equation*}
    \textbf{E}_t = \textbf{X}^{t+1} (\textbf{I} - \textbf{X})^{-1}
\end{equation*}
By taking the matrix norm of both sides, we get
\begin{equation*}
    \|\textbf{E}_t\| \leq \|\textbf{X}^{t+1}\| \cdot \|(\textbf{I} - \textbf{X})^{-1}\|
\end{equation*}
Given that $|\textbf{X}|<1$, it follows that
\begin{equation*}
    \|\textbf{E}_t\| \leq \|\textbf{X}\|^{t+1} \cdot \|(\textbf{I} - \textbf{X})^{-1}\|
\end{equation*}
Thus, we conclude that the relative error for approximating $(1-\textbf{X})^{-1}$ using the first $t$ term of the Neumann series is given by $\|\textbf{X}\|^{t+1}$.


Since $\matrixU^{\mathsf{T}} = 1 - (1 - \gamma) \transmatrix$, to evaluate the approximation error of $\vecty_t = \matrixU^{-T} \indvect{k}$, we set $\mathbf{X} = (1-\gamma) \transmatrix$ with $\gamma = 0.15$. Consequently, $\lVert (1-\gamma) \transmatrix \rVert_\infty = 0.85$. It follows that the relative error of the $t$-term approximation is $0.85^{\,t+1}$. Setting $t = 50$ yields a relative error below $0.0003$.

\section{Omitted algorithms}
\label{appendix:pseudocode}

In this section, we provide pseudocode for Algorithms \ref{alg:project} and \ref{alg:crossgd}. 

\Cref{alg:project} invokes \Cref{alg:root_finding} as a subroutine to compute the optimal Lagrangian dual variable $\lambda^{*}$. The implementation in \Cref{alg:root_finding} differs slightly in form from the procedure described by \citet{adam2022projections}, yet both are mathematically equivalent. For completeness, we now present a formal justification of the correctness of \Cref{alg:root_finding}, establishing that it indeed returns the optimal dual variable.

According to \citet{adam2022projections}, the solution to the projection defined in \Cref{eq:projection_simplex_plus_box} is $s_i^* = \text{clip} (s_i + \lambdaopt, \ell_i, u_i)$, with $\text{clip}(x, a, b)$ defined as
\begin{equation*}
	\text{clip}(x, a, b) = \begin{cases}
		a & \text{if } x \leq a, \\
		x & \text{if } a < x < b, \\
		b & \text{if } x \geq b,
	\end{cases}
\end{equation*}
and $\lambdaopt$ is the solution to the following equation
\begin{equation*}
	h(\lambda) := \sum_{i=1}^{K} \text{clip}(s_i + \lambda, \ell_i, u_i) -1 = 0.
\end{equation*}
Since $h$ is piecewise linear and non-decreasing in $\lambda$, \checkedagtext{we can find~$\lambdaopt$} via a simple method proposed by \citet{adam2022projections}:
\begin{enumerate}[leftmargin=*]
	\item For any $\lambda$, $\clip(s_i, \ell_i, u_i)$ takes value $\ell_i$, $\lambda + s_i$, or $u_i$, thus, we can rewrite $h(\lambda)$ as $h(\lambda) = \act(\lambda) \cdot \lambda + C$ for some constant~$C$, where $\act(\lambda) $ is the number of elements such that $s_i + \lambda$ falls in range $[\ell_i, u_i)$.
	\item Let $R$ be the increasing ranking of $2N$ items consisting of $\vectl - \vects$ and $\vectu - \vects$. For any $\lambda$ in interval $[R_j, R_{j+1})$, the function $\act(\lambda)$ remains constant.
	\item Let $\Total(\lambda) = \sum_{i=1}^N \clip(s_i + \lambda, \ell_i, u_i)$ and assume that $\mathcal{F}_2$ is non-empty. Then there must exist some $j \in \{1, \cdots, 2N\}$ such that $\Total(R_j) \leq 1$ and $\Total(R_{j+1}) > 1$, and we can conclude that $\lambda^*$ lies in range $[R_j, R_{j+1})$. Since $\act(R_j) = \act(\lambda^*)$, it follows that
	$(\lambda^{*} - R_j) \cdot \act(R_j) = 1 - \Total(R_j)$
	and thus we can compute $\lambda^{*}$ as 
	\begin{equation*}
		\lambda^{*} = \frac{1-\Total(R_j)}{\act(R_j)}.
	\end{equation*}
\end{enumerate}


\begin{algorithm}
\caption{\label{alg:crossgd}\crossgd: Projected gradient descent for Group-adapted \phiPageRank and Restricted Group-adapted \phiPageRank}
  \begin{algorithmic}[1]
    \REQUIRE \transmatrix, $\gamma$, \vectv, 
$\groupDist =(\grouppr_1, \ldots, \grouppr_\nogroups)$, group indicator vectors $\indvect{\agroup}$ for $\agroup = \{1, \cdots, \nogroups\}$, learning rate~$\alpha$, iteration exponent~$t_1$ and $t_2$, convergence criterion $\kappa$, maximum number of iterations $N_{\text{iter}}$, modification bound \boundfactor and \absboundfactor (optional).
    \ENSURE Optimized transition matrix $\newtransmatrix$
    \STATE $\newtransmatrix \gets \transmatrix$, ${\iter} \gets 0$, ${\lossprev} \gets \infty$, $\prvector = \frac{1}{n} \indvect{}$
    \WHILE {$|\groupAdapLoss(\newtransmatrix, \gamma) - {\lossprev}| > \kappa$ and $\iter < N_{\text{iter}}$}
      \STATE $\iter \gets \iter + 1$
      \STATE ${\lossprev} \gets \groupAdapLoss(\newtransmatrix, \gamma, \vectv)$
      \FOR {$\ell = 1$ to $\nogroups$}
       \STATE $\prvector \gets \text{iterating } \prvector^{\mathsf{T}} = (1 - \gamma) \prvector^{\mathsf{T}} \newtransmatrix + \frac{\gamma}{|\vertices_{\ell}|} \indvect{\ell}^{\mathsf{T}} \text{ for } t_1 \text{ times}$
        \FOR {$\agroup = 1$ to $\nogroups$}
           \STATE $\vecty_{\agroup} \gets$ $\sum_{i=1}^{t_2} (1-\gamma)^i \newtransmatrix^i \indvect{k}$          
            \STATE $\newtransmatrix \gets \newtransmatrix- \alpha \frac{2(1 - \gamma)}{\nogroups^2} (\indvect{\agroup}^{\mathsf{T}} \prvector_{\ell} - \grouppr_\agroup) \prvector_{\ell} \vecty_{\agroup}$
        \ENDFOR
      \ENDFOR
      \STATE $\newtransmatrix \gets {\project}(\newtransmatrix, \delta, \epsilon)$
    \ENDWHILE
    \RETURN $\newtransmatrix$
  \end{algorithmic}
\end{algorithm}

\begin{algorithm}[t]	\caption{\label{alg:root_finding}\searchlambda: find the optimal dual variable}
	\begin{algorithmic}[1]
		\REQUIRE vector $\vects$, lower bound vector $\vectl$, upper bound vector $\vectu$
		\ENSURE $\lambda^{*}$
            \STATE $K = |\vects|$
            \STATE $\vectlambda \gets$ non-decreasing ranking of $\vectl - \vects$ and $\vectu-\vects$.
            \STATE $\act = 1$, $\Total \gets \sum_j \vectl_j$
            \FOR{$i = 2,\cdots, 2K$}
            \STATE $\Total \gets \Total + \act \times (\vectlambda_i - \vectlambda_{i-1})$
            \IF{$\Total \geq 1$}
            \RETURN $\lambdaopt =  \frac{(1-\Total)}{\act} + \vectlambda_i$
            \ELSE
            \STATE if $\vectlambda_i$ belongs to $\vectl - \vects$ then $\text{Active++}$ else $\text{Active}\texttt{--}$
            \ENDIF
            \ENDFOR
	\end{algorithmic}
\end{algorithm}


\section{Convergence Analysis for minimizing $\loss(\transmatrix)$}
\label{section:convergence_L}
In this section, we first obtain an upper bound for the Lipschitz constant for the loss function, and then prove that the projected gradient descent method converges to a stationary point.

Given restart probability $\gamma$ and restart vector $\vectv$, we can rewrite the loss function as :
\begin{equation*}
    \loss(\transmatrix) = \frac{1}{K} \sum_{k=1}^K \left( g_k(\transmatrix) \right)^2
\end{equation*}
where $g_k(\transmatrix) = \indvect{\agroup}^{\mathsf{T}} \prvector - \phi_k$ and $\prvector = \gamma (\unitary - (1-\gamma) \transmatrix^{\mathsf{T}})^{-1} \vectv.$

\subsection{Gradient of the Loss $\loss(\transmatrix)$}

Though we have already provided the first-gradient analysis of the loss function $\loss(\transmatrix)$, we rewrite it here for ease of analysis. We first compute the gradient of $\loss$ with respect to $\transmatrix_{ij}$:
\begin{equation*}
    \frac{\partial \loss}{\partial \transmatrix_{ij}}
= \frac{2}{K} \sum_{k=1}^K g_k \frac{\partial h_k}{\partial \transmatrix_{ij}}
\end{equation*}
where $ h_k(\transmatrix) = \indvect{\agroup}^{\mathsf{T}} \prvector $.

According to \Cref{eq:gradient_fairness_loss_final},
$\frac{\partial h_k}{\partial \transmatrix} = (1- \gamma )\prvector \indvect{k}^{\mathsf{T} }\invUmatrix$,
therefore,
\begin{equation*}
\frac{\partial h_k}{\partial \transmatrix_{ij}} = (1-\gamma) \prvector_i (\indvect{\agroup}^{\mathsf{T}} \invUmatrix \matr{e}_j)
\end{equation*}
where $\matr{e}_j$ is the standard basis vector (1 at position $j$, zero elsewhere).

\subsection{Second Derivative of the Loss $\loss(P)$}

We now compute the second derivative of the loss:
\begin{align*}
\frac{\partial^2 \loss}{\partial \transmatrix_{ij} \partial \transmatrix_{ab}}
&= \frac{\partial}{\partial \transmatrix_{ab}} \left( \frac{\partial \loss}{\partial \transmatrix_{ij}} \right) \\
&= \frac{\partial}{\partial \transmatrix_{ab}} \left[ \frac{2}{K} \sum_{k=1}^K g_k \frac{\partial h_k}{\partial \transmatrix_{ij}} \right] \\
&= \frac{2}{K} \sum_{k=1}^K \left[
    \frac{\partial g_k}{\partial \transmatrix_{ab}} \frac{\partial h_k}{\partial \transmatrix_{ij}}
    + g_k \frac{\partial^2 h_k}{\partial \transmatrix_{ij} \partial \transmatrix_{ab}}
\right]
\end{align*}
We calculate each term of the second derivative respectively.
\paragraph{First term:}
\[
\frac{\partial g_k}{\partial \transmatrix_{ab}} = \frac{\partial h_k}{\partial \transmatrix_{ab}} = (1-\gamma) \prvector_a (\indvect{\agroup}^{\mathsf{T}} \invUmatrix \matr{e}_b)
\]

\paragraph{Second term:}
\begin{align*}
\frac{\partial^2 h_k}{\partial \transmatrix_{ij} \partial \transmatrix_{ab}}
&= (1-\gamma) \frac{\partial}{\partial \transmatrix_{ab}} \left[ \prvector_i (\indvect{\agroup}^{\mathsf{T}} \invUmatrix \matr{e}_j) \right] \\
&= (1-\gamma) \left[
    \frac{\partial \prvector_i}{\partial \transmatrix_{ab}} (\indvect{\agroup}^{\mathsf{T}} \invUmatrix \matr{e}_j)
    + \prvector_i \frac{\partial}{\partial \transmatrix_{ab}} (\indvect{\agroup}^{\mathsf{T}} \invUmatrix \matr{e}_j)
\right]
\end{align*}
We have:
\[
\frac{\partial \prvector_i}{\partial \transmatrix_{ab}} = (1-\gamma) [\invUmatrix \Ematrix_{ba} \prvector]_i
\]
and
\[
\frac{\partial}{\partial \transmatrix_{ab}} (\indvect{\agroup}^{\mathsf{T}} \invUmatrix \matr{e}_j) = (1-\gamma) \indvect{\agroup}^{\mathsf{T}} \invUmatrix \Ematrix_{ba} \invUmatrix \matr{e}_j
\]
{Combining the two terms, we get}
\begin{align*}
\frac{\partial^2 h_k}{\partial \transmatrix_{ij} \partial \transmatrix_{ab}} &=
(1-\gamma)^2 \Big(
    [\invUmatrix \Ematrix_{ba} \prvector]_i \cdot (\indvect{\agroup}^{\mathsf{T}} \invUmatrix \matr{e}_j)
    + \prvector_i \cdot \indvect{\agroup}^{\mathsf{T}} \invUmatrix \Ematrix_{ba} \invUmatrix \matr{e}_j
\Big)
\end{align*}
Thus, the Hessian entry of $\loss(\transmatrix)$ at coordinate $((ij), (ab))$ is:
\begin{align}
\label{eq:hessian}
    & \frac{\partial^2 \loss}{\partial \transmatrix_{ij} \partial \transmatrix_{ab}}
= \frac{2}{K} \sum_{k=1}^K \Bigg[
        (1-\gamma)^2 \prvector_a \prvector_i (\indvect{\agroup}^{\mathsf{T}} \invUmatrix \matr{e}_b)(\indvect{\agroup}^{\mathsf{T}} \invUmatrix \matr{e}_j)\\ \nonumber
    &+\ g_k(\transmatrix) \cdot (1-\gamma)^2 \Big(
        [\invUmatrix \Ematrix_{ba} \prvector]_i \cdot (\indvect{\agroup}^{\mathsf{T}} \invUmatrix \matr{e}_j)
        + \prvector_i \cdot \indvect{\agroup}^{\mathsf{T}} \invUmatrix \Ematrix_{ba} \invUmatrix \matr{e}_j
    \Big)
\Bigg]
\end{align}

\subsection{\texorpdfstring{Lipschitz Constant for $\tfrac{\partial \loss}{\partial \transmatrix}$}{Lipschitz Constant for dL/d$\transmatrix$}}

Since $\loss(\transmatrix)$ is twice differentiable, the first derivative of $\loss(\transmatrix)$ is Lipschitz smooth with constant $C$ if and only if the operator norm of the Hessian matrix $ \frac{\partial^2 \loss}{\partial \transmatrix_{ij} \partial \transmatrix_{ab}}$ is bounded by $C$ for all $\transmatrix$ in the domain.
In this section, we flat the Hessian matrix and show an upper bound for its operator norm, which serves as an upper bound to its Lipschitz constant.

In order to flatten the Hessian matrix, let $r \in [n^2]$ be the row indexes and $s \in [n^2]$ be the column indexes. Given any $(i,j)$ pair, we can calculate a unique flattened row index $r = (i-1)n + j$ and given any $(a,b)$ pair, we can calculate a unique flattened column index $s = (a-1)n + b$. Conversely, given any pair $(r,s)$ we can calculate the unique corresponding four indices $ i(r), j(r), a(s) \textit{ and } b(s)$ as follows:
\begin{align*}
    &i(r) = \lfloor \frac{r-1}{n}\rfloor + 1, j(r) = r - (i(r) -1)n \\
    &a(s) = \lfloor \frac{s-1}{n}\rfloor + 1, b(s) = s - (a(s) -1)n 
\end{align*}
We can then rewrite \Cref{eq:hessian} as
\begin{align*}
\label{eq:hessian}
     \Hmatrix_{rs}
&= \frac{2}{K} \sum_{k=1}^K \Bigg[
        (1-\gamma)^2 \prvector_{a(s)} \prvector_{i(r)}(\indvect{\agroup}^{\mathsf{T}} \invUmatrix \matr{e}_{b(s)})(\indvect{\agroup}^{\mathsf{T}} \invUmatrix \matr{e}_{j(r)})\\ \nonumber
    &+\ g_k(\transmatrix) \cdot (1-\gamma)^2 \Big(
        [\invUmatrix \Ematrix_{b(s)a(s)} \prvector]_{i(r)} \cdot (\indvect{\agroup}^{\mathsf{T}} \invUmatrix \matr{e}_{j(r)} \\ \nonumber 
        &+ \ g_k(\transmatrix) \cdot (1-\gamma)^2 \prvector_{i(r)} \cdot \indvect{\agroup}^{\mathsf{T}} \invUmatrix \Ematrix_{b(s)a(s)} \invUmatrix \matr{e}_{j(r)}
    \Big)
\Bigg]
\end{align*}
Let 
\begin{align*}
    \Hmatrix_{rs}^{(1)} & = \frac{2}{K} (1-\gamma)^2\sum_{k=1}^K 
         \prvector_{a(s)} \prvector_{i(r)}(\indvect{\agroup}^{\mathsf{T}} \invUmatrix \matr{e}_{b(s)})(\indvect{\agroup}^{\mathsf{T}} \invUmatrix \matr{e}_{j(r)}) \\
        \Hmatrix_{rs}^{(2)} & = \frac{2}{K} (1-\gamma)^2\sum_{k=1}^K 
        \ g_k(\transmatrix) 
        [\invUmatrix \Ematrix_{b(s)a(s)} \prvector]_{i(r)} \cdot (\indvect{\agroup}^{\mathsf{T}} \invUmatrix \matr{e}_{j(r)}) \\    
        \Hmatrix_{rs}^{(3)} & = \frac{2}{K} (1-\gamma)^2\sum_{k=1}^K 
        \ g_k(\transmatrix)  \prvector_{i(r)} \cdot \indvect{\agroup}^{\mathsf{T}} \invUmatrix \Ematrix_{b(s)a(s)} \invUmatrix \matr{e}_{j(r)} 
\end{align*}
Then it holds that
\begin{equation*}
     \Hmatrix_{rs} =  \Hmatrix_{rs}^1 +  \Hmatrix_{rs}^2 +  \Hmatrix_{rs}^3
\end{equation*}
To be able to bound the operator norm of $\Hmatrix$, we bound the operator norm of $\Hmatrix^{(1)},  \Hmatrix^{(2)}, \text{and }  \Hmatrix^{(3)}$ separately.

Before bounding the operator norm of each term, we first proof $\|\invUmatrix\|_2 \leq 1/\gamma$, which will be used multiple times in the following deriviation.
 The proof is the result of three observations: (1) the largest eigenvalue of $\transmatrix^{\mathsf{T}}$ is 1, (2) the smallest eigenvalue of $\matrixU$ is $\gamma$ and (3) the operator norm, $\|\invUmatrix\|_2$, of matrix \invUmatrix is the reciprocal of the smallest absolute eigenvalue of $\matrixU$. We provide details for each observation in the following:
\begin{enumerate}
	\item Let $\indvect{}$ be an all-one vector, i.e. $\indvect{} = (1,1,...,1)^{\mathsf{T}}$. We can see $\indvect{}$ is a eigenvector of $\transmatrix$ since $\transmatrix \textbf{1} = 1$. We also know the largest eigenvalue is bounded by the infinity norm, which is $\| \transmatrix\|_{\infty} = \max_i \sum_{j}|\transmatrix_{ij}| = 1$. Thus, the maximum eigenvalue of $\transmatrix^{\mathsf{T}}$, which is equal to the maximum eigenvalue of $\transmatrix$, is equal to 1.
	
	\item Let $\sigma$ denote each eigenvalue of $\transmatrix^{\mathsf{T}}$, and $\mu$ each eigenvalue of $\matrixU$. Since $\matrixU = \unitary - (1 - \gamma) \transmatrix^{\mathsf{T}}$, it holds that $\mu = 1 - (1 - \gamma) \sigma$. Thus the smallest eigenvalue of $\matrixU$ occurs when $\sigma = 1$, leading to $\mu_{min} = 1 - (1 - \gamma) = \gamma$
	
	\item Combining (1) and (2), we get $\| \invUmatrix\| = \frac{1}{\mu_{min}} = \frac{1}{\gamma}$. 
\end{enumerate}

\subsubsection{Bounding $\|\Hmatrix^{(1)}\|_2$}
Let $\vectw_k$ of a vector of length $n^2$ that depends on group $k$, specifically,  
\begin{equation*}
    \vectw_k[r] = (1 - \gamma) \prvector_{i(r)} (\indvect{k}^{\mathsf{T}} \invUmatrix e_{j(r)})
\end{equation*}
Then we can represent $\Hmatrix^{(1)}$ as a rank 1 matrix as follow:
\begin{equation*}
    \Hmatrix^{(1)} = \frac{2}{K}\sum_{k=1}^{K} \vectw_k \vectw_k^{\mathsf{T}}
\end{equation*}
By the subadditivity of the spectral norm, and the identity $\|\vectw_k \vectw_k^{\mathsf{T}}\|_2 = \|\vectw_k\|_2^2$. We first bound the second norm of $\vectw_k$,  it follows that
\begin{align*}
    \|\vectw_k\|_2^2 &= {(1-\gamma)^2} \sum_{i=1}^{n}\prvector_{i}^2  \sum_{j=1}^{n} (\indvect{k}^{\mathsf{T}} \invUmatrix e_j)^2\\
    & ={(1-\gamma)^2} \sum_{i=1}^{n}\prvector_{i}^2  \|\indvect{k}^{\mathsf{T}} \invUmatrix \|_2^2 \\
    & \stackrel{(a)}{\leq}  \frac{(1-\gamma)^2}{\gamma^2} \|\indvect{k}\|_2^2
\end{align*}
where (a) holds because $\sum_i \prvector_i^2 \leq 1$, and $\|\invUmatrix\| \leq \frac{1}{\gamma}$.
We can then bound $\|\Hmatrix^{(1)}\|_2$ as follow:
\begin{align}
	\label{eq:boundh1}
	\|\Hmatrix^{(1)}\|_2 &= \frac{2}{K}\sum_{k=1}^{K}\|\vectw_k\|_2^2 \\
	& \leq \frac{2 (1 -\gamma)^2}{K \gamma^2} \sum_{k=1}^{K} \|\indvect{k}\|_2^2 = \frac{2 (1 -\gamma)^2}{K \gamma^2} n
\end{align}

\subsubsection{Bounding $\|\Hmatrix^{(2)}\|_2$} 
Let $\Xmatrix \in \mathbb{R}^{n\times n}$ be a matrix, and let $\vectx \in \mathbb{R}^{n^2}$ denote the vector obtained by flattening \Xmatrix.
We can then obtain a vector $\vecty = \Hmatrix^{(2)} \vectx \in \mathbb{R}^{n^2}$, which, after reshaping, is equal to a matrix $\Ymatrix \in \mathbb{R}^{n\times n}$. Specifically, 
\begin{align*}
	 \Ymatrix_{i(r), j(r)} &= \vecty_r  =\sum_{s = 1}^{n^2} \Hmatrix^{(2)}_{r,s} \vectx  \\
	&= \sum_{a(s)=1}^{n} \sum_{b(s)=1}^{n} \Hmatrix^{(2)}_{(i(r),j(r)),(a(s),b(s))} \Xmatrix_{a(s),b(s)} \\
	& \stackrel{(a)}{=} \frac{2(1-\gamma)^2}{K} \sum_{k=1}^{K} g_k \left[ \sum_{a, b} [\invUmatrix \Ematrix_{b a} \prvector]_i \Xmatrix_{a b} \right] (\indvect{k}^{\mathsf{T}} \invUmatrix \vecte_{j})	\\
	&=  \frac{2(1-\gamma)^2}{K} \sum_{k=1}^{K} g_k \left[ \sum_{ b} \left(\invUmatrix \right)_{i,b} \left(\sum_a \prvector_a \Xmatrix_{a,b} \right) \right] (\indvect{k}^{\mathsf{T}} \invUmatrix \vecte_{j})
\end{align*}
where for equality (a) and the following deriviation, we replace $a(s), b(s)$ with $a,b$, and $i(r), j(r)$ with $(i,j)$ for ease of notation whenever the context is clear.

Let $\vectu = \Xmatrix^{\mathsf{T}} \prvector$, $\vectz = \sum_{k=1}^{K} g_k \indvect{k}$, and $\vectw = \matr{A}^{-T} \vectz$, then we can rewrite $\Ymatrix$ as
\begin{equation*}
	\Ymatrix = \frac{2(1-\gamma)^2}{K} (\invUmatrix u) \vectw^{\mathsf{T}}
\end{equation*}
Since  $(\invUmatrix u) \vectw^{\mathsf{T}}$ is a rank-1 matrix, the spectrual norm and the Frobenius norm of $(\invUmatrix u) \vectw^{\mathsf{T}}$ are equal. 
Thus we can bound the operator norm of \Ymatrix as
\begin{align*}
	\|\Ymatrix\|_F & = \frac{2(1-\gamma)^2}{K} \|\invUmatrix \vectu\|_2   \|\vectw\|_2 \\
	&\leq \frac{2(1-\gamma)^2}{K} (\|\invUmatrix\|_2  \|\vectu\|_2)  (\|\invUmatrix\|_2 \|\vectz\|_2)\\
	&\stackrel{(a)}{\leq} \frac{2(1-\gamma)^2}{K \gamma^2} \|\vectu\|_2 \|\vectz\|_2\\
	& \leq \frac{2(1-\gamma)^2}{K \gamma^2} \|\Xmatrix^{\mathsf{T}} \prvector\|_2 \sum_{k=1}^{K} \|\indvect{k}\|_2 |g_k| \\
	& \stackrel{(b)}{\leq}\frac{2(1-\gamma)^2}{K \gamma^2} \|\Xmatrix\|_F \|\prvector\|_2 \sum_{k=1}^{K} \|\indvect{k}\|_2 \\
		& \stackrel{(c)}{\leq}\frac{2(1-\gamma)^2}{K \gamma^2} \|\Xmatrix\|_F  \sqrt{n} 
\end{align*}
where inequality (a) holds because $\|\invUmatrix\|_2 \leq \frac{1}{\gamma}$, (b) holds because $|g_k| \leq 1$ and (c) holds because $\|\prvector\| \leq 1$.

Therefore, it holds that
\begin{align*}
 \|\Hmatrix^{(2)} \vectx \|_2 &= 	 \| \vecty \|_2 = \|\Ymatrix\|_F \leq \frac{2(1-\gamma)^2}{K \gamma^2} \|\Xmatrix\|_F  \sqrt{n} \\
 &= \frac{2(1-\gamma)^2}{K \gamma^2} \| \vectx\|_2  \sqrt{n}
\end{align*}
by dividing both side by $\|\vectx\|_2$ and taking the supremum over $\vectx \neq \matr{0}$, we get
	\begin{equation}
		\label{eq:boundh2}
		\|\Hmatrix^{(2)}\|_2 \leq \frac{2(1-\gamma)^2}{K \gamma^2} \sqrt{n} 
	\end{equation}

\subsubsection{Bounding $\|\Hmatrix^{(3)}\|_2$} We can reorganize terms in $\Hmatrix^{(3)}$ and it holds that
\begin{equation*}
	\Hmatrix^{(3)}_{rs} = \frac{2(1-\gamma)^2}{K} \sum_{k=1}^{K} g_k \prvector_{i(r)} (\invUmatrix)_{a(s),j(r)} (\vectz_k)_{b(s)}
\end{equation*}
where $\vectz_k = \indvect{k}^{\mathsf{T}} \invUmatrix$.

Let $(\Cmatrix_k)_{(i(r), j(r)),(a(s), b(s))} = \prvector_{i(r)} (\invUmatrix)_{a(s), j(r)} (\vectz_k)_{b(s)}$, we can further rewrite $\Hmatrix^{2}$ as follow:
\begin{equation}
	\Hmatrix^{(3)}_{rs} = \frac{2(1-\gamma)^2}{K} \sum_{k=1}^{K} g_k \Cmatrix_k
	\label{eq:H3}
\end{equation}
Furthermore, since the three factors $\prvector_i(r)$, $(\invUmatrix)_{a(s), j(e)}$, $(\vectz_k)_{b(s)}$ depend on disjoint index pairs, $\Cmatrix_k$ can be represented as a triple outer product and reshaped as 
\begin{equation*}
	\Cmatrix_k = \prvector \otimes \invUmatrix \otimes \vectz_k  
\end{equation*}
and thus we can bound the operator norm of $\Cmatrix_k$ as 
\begin{align*}
	\|\Cmatrix_k\|_2  &= \|\prvector\|_2  \|\invUmatrix\|_2 \|\vectz_k\|_2 \\
	& \leq 1 \times \frac{1}{\gamma} \times \|\invUmatrix\|_2\|1_k\|_2 \\
	& \leq  \frac{1}{\gamma^2} \|1_k\|_2
\end{align*}
Therefore, we can bound the operator norm of $\Hmatrix^{(3)}$ as
\begin{align}
	\label{eq:boundh3}
	\|\Hmatrix^{(3)}\|_2 
	& \leq \frac{2K(1-\gamma^2)}{\gamma^2} \sum_{k=1}^{K}\|1_k\|_2\\ \nonumber
	&= \frac{2K(1-\gamma^2)}{\gamma^2} \sum_{k=1}^{K} \sqrt{|\group_k|} \\ \nonumber
	& \stackrel{(a)}{\leq} \frac{2K(1-\gamma^2)}{\gamma^2}  \sqrt{K} \sqrt{ \sum_{k=1}^{K} |\group_k|} \\ \nonumber
	& = \frac{2K(1-\gamma^2)}{\gamma^2}  \sqrt{K} \sqrt{n} \\ \nonumber
	& = \frac{2(1-\gamma^2)}{\gamma^2}  \sqrt{\frac{n}{K}}
\end{align}
where inequality (a) is due to Cauchy-Schwarz inequatlity.

\subsubsection{The Lipschitz constant for the first derivative of $\loss(\transmatrix)$}
Combining \Cref{eq:boundh1}, \Cref{eq:boundh2} and \Cref{eq:boundh3}, it follows that
\begin{equation*}
	\|\Hmatrix\|_2 \leq \|\Hmatrix^{(1)}\|_2 +  \|\Hmatrix^{(2)}\|_2 +  \|\Hmatrix^{(3)}\|_2 \leq \frac{2(1-\gamma)^2}{K \gamma^2} (n+ \sqrt{n} + \sqrt{Kn})
\end{equation*}
Let $C = \frac{2(1-\gamma)^2}{K \gamma^2} (n+ \sqrt{n} + \sqrt{Kn})$, it follows that $\frac{\partial \loss}{\partial \transmatrix}$ is $C$-Lipschitz.

\subsection{Convergence analysis of the projected gradient descent method for minimizing $\loss(\transmatrix)$}
We have proved $\loss(P)$ is a differentiable function and its first derivative is $C$-Lipschitz, since its feasible set $\resfeasiblematrices(\transmatrix) \cap \feasiblematrices(\transmatrix)$ is convex and closed, the projected gradient update with constant step size $0 \leq \alpha \leq 2/C$ converges to a stationary point, which is a local minimum of $\loss(\transmatrix)$ over $\resfeasiblematrices(\transmatrix) \cap \feasiblematrices(\transmatrix)$ \cite{bertsekas1997nonlinear}.

\section{Convergence Analysis for minimizing $\groupAdapLoss(\transmatrix)$}
\label{section:convergence_Lg}

We can derive the Hessian matrix of $\groupAdapLoss(\transmatrix)$ and its Lipschitz constant in a manner analogous to that of $\loss(\transmatrix)$. Specifically, let $\textbf{H}_g$ be the Hessian matrix of $\groupAdapLoss$, then it holds that  
\begin{align}
\label{eq:hessian_adapt}
& \Hmatrix_g[(i,j),(a,b)] = 
     \frac{\partial^2 \groupAdapLoss}{\partial \transmatrix_{ij} \partial \transmatrix_{ab}}  \\ \nonumber
& = \frac{2}{K^2} \sum_{k=1}^K \sum_{l=1}^K \Bigg[
        (1-\gamma)^2 {\prvector_{\ell}}_a {\prvector_{\ell}}_i (\indvect{\agroup}^{\mathsf{T}} \invUmatrix \matr{e}_b)(\indvect{\agroup}^{\mathsf{T}} \invUmatrix \matr{e}_j)\\ \nonumber
    &+\ g_k(\transmatrix) \cdot (1-\gamma)^2 \Big(
        [\invUmatrix \Ematrix_{ba} {\prvector_{\ell}}]_i \cdot (\indvect{\agroup}^{\mathsf{T}} \invUmatrix \matr{e}_j)
        + {\prvector_{\ell}}_i \cdot \indvect{\agroup}^{\mathsf{T}} \invUmatrix \Ematrix_{ba} \invUmatrix \matr{e}_j
    \Big)
\Bigg]
\end{align}
Equation~\eqref{eq:hessian_adapt} follows directly from \Cref{eq:hessian} by replacing $\prvector$ with $\prvector_{\ell}$ throughout, introducing an additional summation over $l \in {1, \ldots, \nogroups}$, and dividing by the extra term $K$.
Let $\Hmatrix_g^k$ denote the inner sum of $\Hmatrix_g$, that is,
\begin{align*}
   & \Hmatrix_g^k = \frac{2}{K} \sum_{l=1}^K \Bigg[
        (1-\gamma)^2 {\prvector_{\ell}}_a {\prvector_{\ell}}_i (\indvect{\agroup}^{\mathsf{T}} \invUmatrix \matr{e}_b)(\indvect{\agroup}^{\mathsf{T}} \invUmatrix \matr{e}_j)\\ \nonumber
    &+\ g_k(\transmatrix) \cdot (1-\gamma)^2 \Big(
        [\invUmatrix \Ematrix_{ba} {\prvector_{\ell}}]_i \cdot (\indvect{\agroup}^{\mathsf{T}} \invUmatrix \matr{e}_j)
        + {\prvector_{\ell}}_i \cdot \indvect{\agroup}^{\mathsf{T}} \invUmatrix \Ematrix_{ba} \invUmatrix \matr{e}_j
    \Big)
\Bigg]
\end{align*}
It is straightforward to see that the derivation of an upper bound on the operator norm of $\Hmatrix$ remains valid for $\Hmatrix_g^k$ when $\prvector$ is replaced by $\prvector_{\ell}$. That is, $\Hmatrix$ and $\Hmatrix_g^k$ share the same operator norm upper bound $C$. 
Given the subadditivity property of the operator norm, it holds that
\begin{equation*}
    \|\Hmatrix_g\|_2 \leq \frac{1}{\nogroups} \sum_{k=1}^{K} \|\Hmatrix_g^k\|_2 = \|\Hmatrix\|_2 = C
\end{equation*}
Therefore, we have established that the operator norm of $\groupAdapLoss$ is bounded by $C$, which is equivalent to stating that the Lipschitz constant of its gradient is also bounded by $C$. Consequently, projected gradient descent converges to a local minimum when using a constant step size $0 \leq \alpha \leq 2/C$.

\section{Experimental settings}

\subsection{Datasets}
\label{appendix:datasets}

\noindent 
\book \footnote{\url{http://www-personal.umich.edu/~mejn/netdata/}} This is an undirected network of political books sold on Amazon. Each edge represents frequent co-purchases of two books. The books are categorized into three groups: liberal, neutral, and conservative. Since the network is undirected, each edge is converted into two directed edges. 

\smallskip
\noindent 
\blogs \citep{adamic_political_2005}. A directed network of political blogs, where each edge represents a hyperlink between two blogs. The vertices are divided into two groups: liberal and conservative blogs.

\smallskip
\noindent 
\mind \citep{wu2020mind} A dataset built from user impression logs on Microsoft News. Each log captures a user's impressions of current news articles, along with their reading history. News articles are labeled, and we construct a directed graph from these logs, filtering out news articles with a frequency of fewer than 50 occurrences. A directed edge is created from a historically read news article to a current one if both appear in the same user's log. The two vertex groups correspond to lifestyle-related and society-related news\fnlabels.

\smallskip
\noindent 
\twitter \citep{nr-sigkdd16}. \Amatrix directed network of political retweets, where each edge represents a retweet from one user to another. The vertex groups are left-leaning and right-leaning users.

\smallskip
\noindent 
\slashdot \citep{leskovec_community_2009}. \Amatrix directed network representing friend/foe relationships between users on Slashdot. Since this dataset lacks group labels, we apply the METIS graph partitioning algorithm \citep{karypis_metis_1997} to divide the vertices into groups.

\subsection{Baselines}
\label{appendix:baselines}

We present here a description of the baselines we use in our evaluation.
In our experiments, 
\newtransmatrix represents the revised transition matrix, and 
\newrestartvec represents the revised restart vector for both our algorithms and all the baselines.

\smallskip
\noindent 
\lfpru\citep{tsioutsiouliklis_fairness-aware_2021}. 
\Amatrix residual-based algorithm that revises both the restart vector and the transition matrix to achieve perfect $\groupDist$-fairness in \pagerank scores. The intuition is that each node distributes a portion of its PageRank to the under-represented groups, and distributes the remaining portion to its neighbors uniformly. 

The revised transition matrix is $\newtransmatrix = \transmatrix_L + \mathbf{R}$, where:
\begin{equation}
    \transmatrix_L =  
    \left\{\begin{matrix}
        \frac{1-\grouppr_1}{out_2(i)}, &  \text{if } (i,j)\in \edges \text{ and } i \in L_1\\
         \frac{\grouppr_1}{out_1(i)}, &  \text{if } (i,j)\in \edges \text{ and } i \in L_2\\
         0 & \text{otherwise} \\
    \end{matrix}\right.
    \end{equation}
and $out_{\agroup}(i)$ represents the number of out-neighbors of vertex $i$ that belongs to group \agroup, and $L_\agroup$ denotes the set of vertices whose portion of neighbors that belong to group \agroup is smaller than the fairness target $\grouppr_\agroup$. $\mathbf{R}$ is given by
\begin{equation}
    \mathbf{R} = \delta_1 \mathbf{x}^{\mathsf{T}} + \delta_2 \mathbf{y}^{\mathsf{T}}
\end{equation}
where $\delta_1[i] = \grouppr_1 -\frac{(1-\grouppr_1)out_1(i)}{out_2(i)}$, $\delta_2[i] = (1-\grouppr_1) -\frac{\grouppr_1 out_2(i)}{out_1(i)}$. $\mathbf{x}$ and $\mathbf{y}$ are two indicator vectors where $\mathbf{x}(i) = 1 \text{ if }  i\in \group_1$ and $\mathbf{y}(i) = 1 \text{ if } i\in \group_2$.

\smallskip
\noindent 
\lfprn \citep{tsioutsiouliklis_fairness-aware_2021}. 
\Amatrix neighborhood-based algorithm that achieves perfect $\groupDist$-fairness in \pagerank scores. The revised transition matrix is given by $\newtransmatrix = \sum_{\agroup =1}^{2}\grouppr_\agroup \transmatrix_\agroup$, where 
\begin{equation}
    \transmatrix_\agroup =  \left\{\begin{matrix}
        \frac{1}{out_\agroup(i)}, &  \text{if } (i,j) \in E \text{ and } j \in \group_k \\
         \frac{1}{|\group_k|}, & \text{if } out_\agroup(i) = 0 \text{ and } j \in \group_k \\
         0 &  \text{otherwise} \\
\end{matrix}\right.
\end{equation}

\smallskip
\noindent 
\fairwalk \citep{rahman_fairwalk_2019}. 
This algorithm reweights the edges such that a random walker has an equal probability of visiting each reachable group and an equal probability of visiting each vertex within each reachable group. \fairwalk does not change the restart vector, and the revised transition matrix is given by:
\begin{equation}
\label{eq:fairwalk}
    \newtransmatrix[i,j] = \frac{\grouppr_{\labelof_j} \transmatrix[i,j]}{\sum_{m \in \group_{\labelof_j}}P[i,m] \sum_{\agroup =1}^{\nogroups} I_{\agroup}(i)\grouppr_\agroup} ,
\end{equation}
where $I_\agroup(i) = 1$ if vertex $i$ has a out-neighbor that belongs to group $\agroup$ and 0 otherwise.

\smallskip
\noindent 
\crosswalk \citep{khajehnejad_crosswalk_2022}. 
This algorithm extends \fairwalk by taking into account the distance of each vertex to other groups in the random walk. \crosswalk upweights edges that (1) are closer to the group boundaries or (2) connect different groups in the graph. This algorithm does not change the restart vector, and we refer readers to Algorithm 1 in the paper for the revised transition matrix.

\section{Additional experimental results} 
\label{appendix:more-experiments}

\subsection{Ranking Coefficient for the group-wise \pagerank weights}

In Tables \ref{table:ranking_blog} and \ref{table:ranking_slashdot}, we report the weighted average Spearman’s rank correlation coefficient, $\overline{\rho}$, for all methods on the \blogs and \slashdot datasets, respectively. Results for the \book and \twitter datasets are omitted, as the average coefficients are close to zero across all methods, indicating that none of the methods can preserve the original \pagerank rankings within groups.

On the \blogs dataset, \crossgdrone achieves the highest correlation coefficient, followed by \lfprn and \lfpru. \fairgdrtwo and \crossgd obtain the lowest values, while the remaining methods yield comparable performance. 
On the \slashdot dataset, algorithm \crossgdrone again achieves the highest value, followed by \crossgdrtwo. Algorithms \crossgd, \fairgd, and its variants \fairgdrone and \fairgdrtwo produce comparable results, whereas \fairwalk and \crosswalk perform the worst.

\subsection{Ranking Coefficient for the edge weights}

In Tables \ref{table:edge_ranking_book} to \ref{table:edge_ranking_mind}, we present the average Spearman’s rank correlation coefficient, $\tilde{\rho}$, for each method on the \book, \blogs and 
\mind dataset. For each node, we compute the Spearman’s coefficient between the rankings of its original and updated outgoing edge weights, and then report the average coefficient across all nodes.
We omit the results for the datasets \twitter and \slashdot, as the average coefficients are all approximately in the range $[0.999–1)$ across all methods.

\fairwalk performs consistently well across all three datasets, achieving the best results on the \book dataset and second-best results on the remaining datasets. \crossgdrone and \crossgd($0.5$,$0.1$) attain the second-best performance on the \book and \blogs datasets, and the third-best on the \mind dataset. \fairgd achieves the optimal outcome on the \mind dataset.

For the two-group datasets, \lfpru and \lfprn perform noticeably worse than the other methods. This outcome is expected, as both methods substantially modify the underlying graph structure.

In conclusion, \fairwalk best preserves the edge weight ranking across datasets, while \crossgdrone and \crossgdrtwo also perform very well.

\begin{table*}[t]
\centering
\caption{\label{table:ranking_blog}
Average Spearman's coefficient $\overline{\rho}$ between the \pagerank weight of each method for the \blogs dataset}
\vspace{-2mm}
\begin{small}
\begin{tabular}{c|cccccccccc}
\hline
$\phi$ & \fairwalk & \crosswalk &  \lfpru & \lfprn & \fairgd & $\fairgd(0.1, 0.1)$ & $\fairgd(0.5, 0.1)$ & \crossgd & $\crossgd(0.1, 0.1)$ & $\crossgd(0.5, 0.1)$ \\
\hline
0.1 & \underline{0.573} & 0.506 & 0.558 & 0.547 & 0.437 & 0.423 & 0.437 & 0.423 & \textbf{0.594} & 0.561 \\
0.2 & 0.579 & 0.527 & 0.552 & \textbf{0.591} & 0.483 & 0.488 & 0.515 & 0.446 & \underline{0.587} & 0.568 \\
0.3 & 0.568 & 0.497 & 0.577 & \underline{0.593} & 0.534 & 0.553 & 0.542 & 0.433 & \textbf{0.598} & 0.553 \\
0.4 & \underline{0.602} & 0.539 & 0.566 & 0.597 & 0.560 & 0.555 & 0.540 & 0.442 & \textbf{0.616} & 0.551 \\
0.5 & \underline{0.590} & 0.513 & 0.588 & 0.577 & 0.562 & 0.564 & 0.577 & 0.451 & \textbf{0.609} & 0.557 \\
0.6 & 0.576 & 0.490 & 0.599 & 0.579 & 0.557 & 0.569 & 0.570 & \textbf{0.801} & \underline{0.614} & 0.545 \\
0.7 & 0.588 & 0.511 & 0.604 & 0.567 & 0.586 & 0.574 & 0.573 & \textbf{0.754} & \underline{0.627} & 0.570 \\
0.8 & \underline{0.612} & 0.502 & 0.602 & \textbf{0.613} & 0.510 & 0.489 & 0.493 & 0.492 & 0.604 & 0.575 \\
0.9 & 0.584 & 0.498 & 0.606 & \underline{0.617} & 0.516 & 0.506 & 0.501 & 0.503 & \textbf{0.620} & 0.555 \\
\hline
\end{tabular}
\end{small}
\end{table*}

\begin{table*}[t]
\centering
\caption{
Average Spearman's coefficient $\overline{\rho}$ between the \pagerank weight of each method for the \slashdot dataset}
\vspace{-2mm}
\begin{small}
\begin{tabular}{c|cccccccccc}
\hline
$\phi$ & \fairwalk & \crosswalk &  \fairgd & $\fairgd(0.1, 0.1)$ & $\fairgd(0.5, 0.1)$ & \crossgd & $\crossgd(0.1, 0.1)$ & $\crossgd(0.5, 0.1)$ \\
\hline
0.1 & 0.123 & 0.124 & 0.221 & 0.215 & 0.217 & 0.212 & \textbf{0.295} & \underline{0.244} \\
0.2 & 0.140 & 0.131 & 0.219 & 0.217 & 0.216 & 0.213 & \textbf{0.312} & \underline{0.251} \\
0.3 & 0.146 & 0.135 & {0.245} & 0.239 & 0.241 & 0.238 & \textbf{0.310} & \underline{0.251} \\
0.4 & 0.135 & 0.143 & 0.229 & 0.231 & 0.235 & {0.243} & \textbf{0.308} & \underline{0.271} \\
0.5 & 0.121 & 0.120 & 0.218 & 0.212 & 0.213 & {0.223} & \textbf{0.296} & \underline{0.261} \\
0.6 & 0.131 & 0.119 & 0.221 & 0.228 & {0.228} & 0.211 & \textbf{0.287} & \underline{0.254} \\
0.7 & 0.129 & 0.117 & 0.226 & 0.229 & {0.232} & 0.212 & \textbf{0.282} & \underline{0.248} \\
0.8 & 0.128 & 0.117 & 0.235 & 0.230 & {0.235} & 0.224 & \textbf{0.280} & \underline{0.250 }\\
0.9 & 0.125 & 0.122 & 0.230 & 0.224 & {0.229} & 0.224 & \textbf{0.271} & \underline{0.254} \\
\hline
\end{tabular}
\end{small}
\label{table:ranking_slashdot}
\end{table*}

\begin{table*}[t]
\centering
\caption{
Average Spearman's coefficient $\tilde{\rho}$ between edge weight rankings for the \book dataset}
\vspace{-2mm}
\begin{small}
\begin{tabular}{c|cccccccccc}
\hline
$\phi$ & \fairwalk & \crosswalk &  \fairgd & $\fairgd(0.1, 0.1)$ & $\fairgd(0.5, 0.1)$ & \crossgd & $\crossgd(0.1, 0.1)$ & $\crossgd(0.5, 0.1)$ \\
\hline
0.1 & \textbf{0.989} & 0.974 & 0.897 & 0.919 & 0.889 & 0.853 & 0.975 & \underline{0.976} \\
0.2 & \textbf{0.989} & 0.975 & \underline{0.976} & 0.919 & 0.940 & 0.855 & \underline{0.976} & 0.975 \\
0.3 & \textbf{0.989} & 0.975 & 0.963 & 0.946 & 0.976 & 0.869 & \underline{0.977 }& \underline{0.977} \\
0.4 & \textbf{0.989} & 0.974 & 0.969 & 0.959 & 0.964 & 0.871 & \underline{0.977} & \underline{0.977 }\\
0.5 & \textbf{0.989} & 0.973 & 0.965 & 0.961 & 0.972 & 0.862 & \underline{0.978} & 0.976 \\
0.6 & \textbf{0.988} & 0.976 & 0.973 & 0.945 & 0.950 & 0.879 & \underline{0.979} & 0.977 \\
0.7 & \textbf{0.988} & 0.975 & 0.942 & 0.943 & 0.932 & 0.865 & \underline{ 0.977} & \underline{0.977} \\
0.8 & \textbf{0.988} & 0.974 & 0.921 & 0.939 & 0.922 & 0.865 & 0.976 & \underline{0.977} \\
0.9 & \textbf{0.988} & 0.974 & 0.907 & 0.938 & 0.920 & 0.909 & 0.977 & \underline{0.977} \\
\hline
\end{tabular}
\end{small}
\label{table:edge_ranking_book}
\end{table*}

\begin{table*}[t]
\centering
\caption{\label{table:edge_ranking_blog}
Average Spearman's coefficient $\tilde{\rho}$ between edge weight rankings on the \blogs dataset}
\vspace{-2mm}
\begin{small}
\begin{tabular}{c|cccccccccc}
\hline
$\phi$ & \fairwalk & \crosswalk &  \lfpru & \lfprn & \fairgd & $\fairgd(0.1, 0.1)$ & $\fairgd(0.5, 0.1)$ & \crossgd & $\crossgd(0.1, 0.1)$ & $\crossgd(0.5, 0.1)$ \\
\hline
0.1 & \textbf{0.997} & 0.991 & -0.249 & 0.282 & 0.987 & 0.987 & 0.987 & 0.984 & \underline{0.996} & \underline{0.996} \\
0.2 & \textbf{0.997}& 0.991 & -0.142 & 0.284 & 0.992 & 0.991 & 0.992 & 0.984 & 0.995 & \underline{0.996} \\
0.3 & \textbf{0.997} & 0.991 & -0.107 & 0.284 & 0.995 & 0.995 & 0.995 & 0.986 & \underline{0.996} & \underline{0.996}\\
0.4 & \textbf{0.997} & 0.991 & -0.074 & 0.284 & 0.995 & 0.995 & 0.995 & 0.986 & \underline{0.996} & \underline{0.996} \\
0.5 & \textbf{0.997} & 0.991 & 0.157  & 0.486 & 0.995 & 0.995 & 0.995 & 0.988 & \underline{0.996} & \underline{0.996} \\
0.6 & \textbf{0.997} & 0.991 & 0.161  & 0.486 & 0.995 & 0.995 & 0.995 & 0.995 & \underline{0.996} & \underline{0.996} \\
0.7 & \textbf{0.997} & 0.991 & 0.186  & 0.486 & 0.995 & 0.995 & 0.995 & 0.995 & \underline{0.996} & \underline{0.996} \\
0.8 & \textbf{0.997} & 0.991 & 0.226  & 0.486 & 0.991 & 0.990 & 0.990 & 0.984 & 0.995 & \underline{0.996} \\
0.9 & \textbf{0.997} & 0.991 & 0.290  & 0.486 & 0.987 & 0.988 & 0.987 & 0.984 & 0.995 & \underline{0.996} \\
\hline
\end{tabular}
\end{small}
\end{table*}

\begin{table*}[t]
\centering
\caption{\label{table:edge_ranking_mind}
Average Spearman's coefficient $\tilde{\rho}$ between edge weight rankings on the \mind dataset.}
\vspace{-2mm}
\begin{small}
\begin{tabular}{c|cccccccccc}
\hline
$\phi$ & \fairwalk & \crosswalk &  \lfpru & \lfprn & \fairgd & $\fairgd(0.1, 0.1)$ & $\fairgd(0.5, 0.1)$ & \crossgd & $\crossgd(0.1, 0.1)$ & $\crossgd(0.5, 0.1)$ \\
\hline
0.1 & \underline{0.999} & 0.998 & 0.318 & 0.842 & \textbf{$\approx 1$} & \textbf{$\approx 1$} & \textbf{$\approx 1$} & 0.998 & 0.998 & \underline{0.999} \\
0.2 & \underline{0.999} & 0.998 & 0.306 & 0.842 & \textbf{$\approx 1$} & \textbf{$\approx 1$} & \textbf{$\approx 1$} & 0.998 & \underline{0.999} & 0.998 \\
0.3 & \underline{0.999} & 0.998 & 0.306 & 0.842 & \textbf{$\approx 1$} & \textbf{$\approx 1$} & \textbf{$\approx 1$} & 0.998 & \underline{0.999} & 0.998 \\
0.4 & \underline{0.999} & 0.998 & 0.298 & 0.842 & \textbf{$\approx 1$} & \textbf{$\approx 1$} & \textbf{$\approx 1$} & 0.998 & \underline{0.999} & \underline{0.999} \\
0.5 & \underline{0.999} & 0.998 & 0.298 & 0.842 & \textbf{$\approx 1$} & \underline{0.999} & \textbf{$\approx 1$} & 0.998 & 0.998 & 0.998 \\
0.6 & \underline{0.999} & 0.998 & 0.298 & 0.842 & \textbf{$\approx 1$} & \textbf{$\approx 1$} & \underline{0.999} & 0.998 & 0.998 & 0.998 \\
0.7 & 0.999 & 0.998 & 0.295 & 0.842 & \textbf{$\approx 1$} & \textbf{$\approx 1$} & \underline{0.999} & 0.998 & 0.998 & 0.998 \\
0.8 & 0.999 & 0.998 & 0.295 & 0.842 & \textbf{$\approx 1$} & \underline{0.999} & \underline{0.999} & 0.997 & 0.998 & 0.998 \\
0.9 & \underline{0.999} & 0.998 & 0.294 & 0.842 & \textbf{$\approx 1$} & \underline{0.999} & \underline{0.999} & 0.996 & 0.998 & 0.998 \\
\hline
\end{tabular}
\end{small}
\end{table*}

\end{document}